\documentclass[aps,prl,notitlepage,superscriptaddress,twocolumn]{revtex4-1}
\usepackage[utf8]{inputenc}
\usepackage[english]{babel}
\usepackage{amsmath}
\usepackage{amsfonts}
\usepackage{amssymb}
\usepackage{listings}
\usepackage{lipsum}
\usepackage{multirow}
\usepackage{datetime}
\usepackage{graphicx}
\usepackage{mathtools}
\usepackage{mathrsfs}
\usepackage{dcolumn}
\usepackage{multirow}
\usepackage{orcidlink}
\usepackage{color}   
\usepackage{hyperref}
\hypersetup{
    colorlinks=true, 
    pdfborder = {0 0 0.5 [3 3]},
    anchorcolor=black,
    citecolor=blue,
    linktoc=all,    
    linktocpage=true,
    linkcolor=red,
	urlcolor=blue
}

\newcommand{\red}[1]{{{#1}}}

\begin{document}

\title{Weakly turbulent saturation of the nonlinear scalar ergoregion instability}
\author{Nils Siemonsen\orcidlink{0000-0001-5664-3521}}
\email[]{nils.siemonsen@princeton.edu}
\affiliation{Princeton Gravity Initiative, Princeton University, Princeton, NJ 08544, USA}
\affiliation{Department of Physics, Princeton University, Princeton, NJ 08544, USA}

\date{\today}

\begin{abstract} 
We perform time-domain evolutions of the ergoregion instability on a horizonless spinning ultracompact spacetime in scalar theories with potential-type and derivative self-interactions mimicking the nonlinear structure of the Einstein equations. We find that the instability saturates by triggering a weakly turbulent direct cascade, which transfers energy from the most unstable and large-scale modes to small scales. The cascade's nonlinear timescales of each mode are orders of magnitude shorter than the corresponding linear e-folding times. Through this mechanism, the counter-rotating stable light ring is filled with a spectrum of higher-order azimuthal modes forming a ring-like shape. Thereby we demonstrate that turbulent processes are likely also important during the fully gravitational saturation of the instability, leaving imprints in the gravitational wave emission.
\end{abstract}

\maketitle

\textbf{Introduction --} \red{Sufficiently compact horizonless spinning objects in general relativity can exhibit two relativistic features setting them apart from vacuum black holes and less compact objects: the existence of stable light rings and ergoregions disconnected from any horizon. On the former, null geodesics (or high frequency gravitational waves) can be stably trapped around the object. Is the object rapidly spinning, an ergoregion encloses this counter-rotating stable light ring~\cite{Cardoso:2014sna}; that is, null geodesics on this light ring (i.e., with angular momentum opposite to that of the object) are forced to orbit with the object. At the linear level, the presence of an ergoregion has been shown to lead to the so-called ergoregion instability~\cite{Friedman:1978,Vilenkin:1978uc,Moschidis:2016zjy}. Physically, there exist massless field configurations with negative energy inside the ergoregion, which radiate positive energy flux to infinity. By energy conservation, this energy flux is balanced by an unbounded growing amount of negative energy inside the ergoregion, signaling an unstable process. On the other hand, stable light rings have been conjectured to trigger a nonlinear instability mechanism due to accumulation of interactions between long-lived massless modes \cite{Keir:2014oka,Cardoso:2014sna}. Using numerical relativity such a nonlinear instability was claimed to exist in spinning boson star spacetimes with stable light rings in Ref.~\cite{Cunha:2022gde}; however, thus far these results could not be reproduced \cite{Evstafyeva:2025mvx,Siemonsen:2025wib} (see also Refs.~\cite{Marks:2025jpt,Siemonsen:2024snb}). In the context of rapidly spinning ultra compact spacetimes, these two phenomena are intrinsically linked as the counter-rotating stable light ring is enclosed by the ergoregion.

In general relativity, stable trapping of null geodesics can lead to a turbulent direct cascade---a nonlinear mechanism connecting linear modes and transporting energy from large scales to small scales. For instance, the perfectly trapping Anti-de Sitter spacetime was found to exhibit gravitational turbulence, ultimately leading to black hole formation at small scales \cite{Bizon:2011gg,Jalmuzna:2011qw,Buchel:2012uh,Balasubramanian:2014cja,Craps:2014vaa,Bizon:2015pfa,Evnin:2021buq} (such turbulent mechanisms have also been studied in other contexts~\cite{Galtier:2017mve,Galtier:2021ovg,Benomio:2018ivy,Bhattacharyya:2007vjd,Carrasco:2012nf,Adams:2013vsa,Green:2013zba,Krynicki:2025fzi,Gay:2024kay,Liang:2025lek,Andrade:2019rpn,Yang:2014tla,Iuliano:2024ogr,Figueras:2023ihz,Ma:2025rnv,Detweiler:1977gy,Hod:2008zz,Yang:2012pj,Cardoso:2013pza,Holzegel:2013kna}). A step away from this nonlinear gravitational setting, the dynamics of a massless scalar testfield with quartic self-interactions trapped in a stable light ring were studied in Refs.~\cite{Benomio:2024lev,Redondo-Yuste:2025hlv}. This scalar field was found to exhibit wave turbulence for sufficiently large initial data, triggered by the nonlinear interactions between long-lived linear modes. While such turbulent phenomena are ubiquitous in systems described by a nonlinear Schrödinger-type equation \cite{zakharov2012kolmogorov,Galtier_2022,nazarenko}, the dispersive nature of massless fields acts against accumulation of nonlinear interactions, unless in confining settings.

Both stable light rings and ergoregions are universal features of large classes of exotic compact objects and black hole mimickers~\cite{Cardoso:2019rvt,Bambi:2025wjx,Carballo-Rubio:2025fnc}, arising in quantum gravity models and beyond the Standard Model physics. In fact, proposed methods of searching for, or constraining the existence of, these objects using gravitational waves hinge on accurate understanding of the nonlinear evolution of such spacetimes~\cite{Barausse:2018vdb,Mastrogiovanni:2025ixe,Fan:2017cfw}. Thus, the presence and dynamics of turbulent processes associated with stable light rings, as well as the nonlinear development of the ergoregion instability, have strong implications for the existence and evolution of exotic compact objects and black hole mimickers; as well as the theories they arise from. Despite this, little is known about the nonlinear development of the ergoregion instability beyond simplistic expectations, the presence of a turbulent mechanism during its nonlinear evolution, and the impact of such a mechanism on the subsequent dynamics and final state of the system.

In this work we investigate the nonlinear saturation of the ergoregion instability for the first time. As a model of the full Einstein equations, we focus on a massless scalar testfield with derivative and potential-type self-interactions propagating on a (horizonless) spinning boson star background with ergoregion. Our key results are (i) that the fastest-growing ergoregion unstable state triggeres a turbulent direct cascade in the nonlinear regime, (ii) this turbulent cascade populates modes more and more localized on the stable light ring, and (iii) once triggered the nonlinear timescales associated with this process are much shorter than the linear instability timescales.} Finally, we speculate about the late-time dynamics of the system, as well as discuss implications for observational searches for black hole mimickers. We set $G=c=1$ throughout.

\red{\textbf{Methods --} In potential astrophysical ultra compact objects, the ergoregion instability is driven by gravitational wave emission and nonlinear effects are governed by the Einstein equations. As a model of this, we} solve the nonlinear complex scalar field equation
\begin{align}
\square_g\Psi=\kappa \Psi |\Psi|^2+\alpha\Psi^* g^{\mu\nu}\partial_\mu\Psi\partial_\nu\Psi,
\label{eq:KGeq}
\end{align}
\red{where} ${}^*$ denotes complex conjugation, and $g_{\mu\nu}$ being the background's metric. Eq.~\eqref{eq:KGeq} mimics the nonlinear derivative structure of the Einstein equations, $\partial^2 g\sim g(\partial g)^2$, \red{and hence, acts as a proxy of full general relativity.} We refer to the $\sim\kappa \Psi |\Psi|^2$ piece as potential-type self-interactions~\footnote{\red{The derivative self-interactions descend from the Lagrangian $\mathcal{L}=(1-\alpha|\Psi|^2)|\partial\Psi|^2$ at leading order in $\alpha|\Psi|^2\ll 1$.}}. \red{In the linear regime, i.e., $\alpha=\kappa=0$, scalar field configurations subject to the ergoregion instability in axisymmetric spacetimes are non-axisymmetric with azimuthal index $m$ and complex frequency $\omega_m$. Let $m>0$ be the family of unstable modes, then the growth timescales of the energy of these modes was shown to increase as $\tau^{\rm EI}_m =[2 \ \text{Im}(\omega_m)]^{-1}\sim e^{a m}$, with $a>0$, in the eikonal limit $m\rightarrow\infty$ \cite{Comins:1978}. In this limit, such massless modes approach the dynamics of stably trapped null geodesics in the counter-rotating light ring contained within the ergoregion \cite{Comins:1978,Siemonsen:2025wib}; in particular, the unstable mode's pattern speed $\Sigma=\text{Re}(\omega_m)/m$ approaches those geodesics' orbital frequency $\omega_-$: $\Sigma\rightarrow \omega_-$ (see e.g., Ref.~\cite{Siemonsen:2025wib}). Intuitively, the decay rates of the longest lived massless modes, trapped in the counter-rotating light ring, turn negative as soon as the spacetime carries sufficient angular momentum to spawn an ergoregion.}

The ultra compact object spacetime is chosen to be that of a (stationary and axisymmetric) spinning boson star. In the slow-rotation approximation \cite{Hartle1967}, the metric takes the form
\begin{align}
ds^2= f dt^2 + & \frac{l}{f} \{dr^2 + r^2 d\theta^2+ r^2 \mathrm{sin}^2 \theta (d\varphi - \Omega dt)^2 \},
\label{eq:metric}
\end{align}
where $f,l,\Omega$ (functions of $r$ only) are obtained numerically using methods outlined in Refs.~\cite{Kleihaus:2005me,Siemonsen:2020hcg}. In the ergoregion instability context, this approximation was used, for example, in Refs.~\cite{Comins:1978,Cardoso:2007az,Siemonsen:2025wib} and captures all relevant features; see Ref.~\cite{Siemonsen:2025wib} for a detailed study of its accuracy. We expand the scalar field as $\Psi=r^{-1}\sum_I \phi_{\ell m}(t,r)Y_{\ell m}(\theta,\varphi)$, where $Y_{\ell m}$ are regular spherical harmonics, as well as define the index set $I=\{(\ell,m)\in \mathbb{Z}^2|\ell\geq 0, |m|\leq \ell\}$. With this, eq.~\eqref{eq:KGeq} turns into a tower of coupled radial wave equations:
\begin{align}
(\mathcal{D}-V_{\bar{\ell}\bar{m}})\phi_{\bar{\ell}\bar{m}}= \sum_{I,I',I''} N^{\bar{m}m m' m''}_{\bar{\ell}\ell\ell'\ell''}(\Psi).
\label{eq:scalar_equations}
\end{align}
All modes, $\phi_{\bar{\ell} \bar{m}}(t,r)$, are coupled through the coefficients $N^{\bar{m}m m' m''}_{\bar{\ell}\ell\ell'\ell''}(\Psi)$ containing the scalar self-interactions (presented explicitly in~\cite{sm}). The differential operator reads
\begin{align}
\mathcal{D}\phi_{\ell m}=-\partial_t^2\phi_{\ell m}+\frac{f^2}{l^{3/2}}\partial_r\left(l^{1/2}\partial_r\phi_{\ell m}\right) - 2im\Omega\partial_t\phi_{\ell m},
\end{align}
whereas the potential is
\begin{align}
V_{\ell m}=\frac{f^2}{l}\left(\frac{\ell(\ell+1)}{r^2}+\frac{\partial_rl}{2lr}-\frac{m^2\Omega^2l}{f^2}\right).
\end{align}

\begin{figure}
\includegraphics[width=0.99\linewidth]{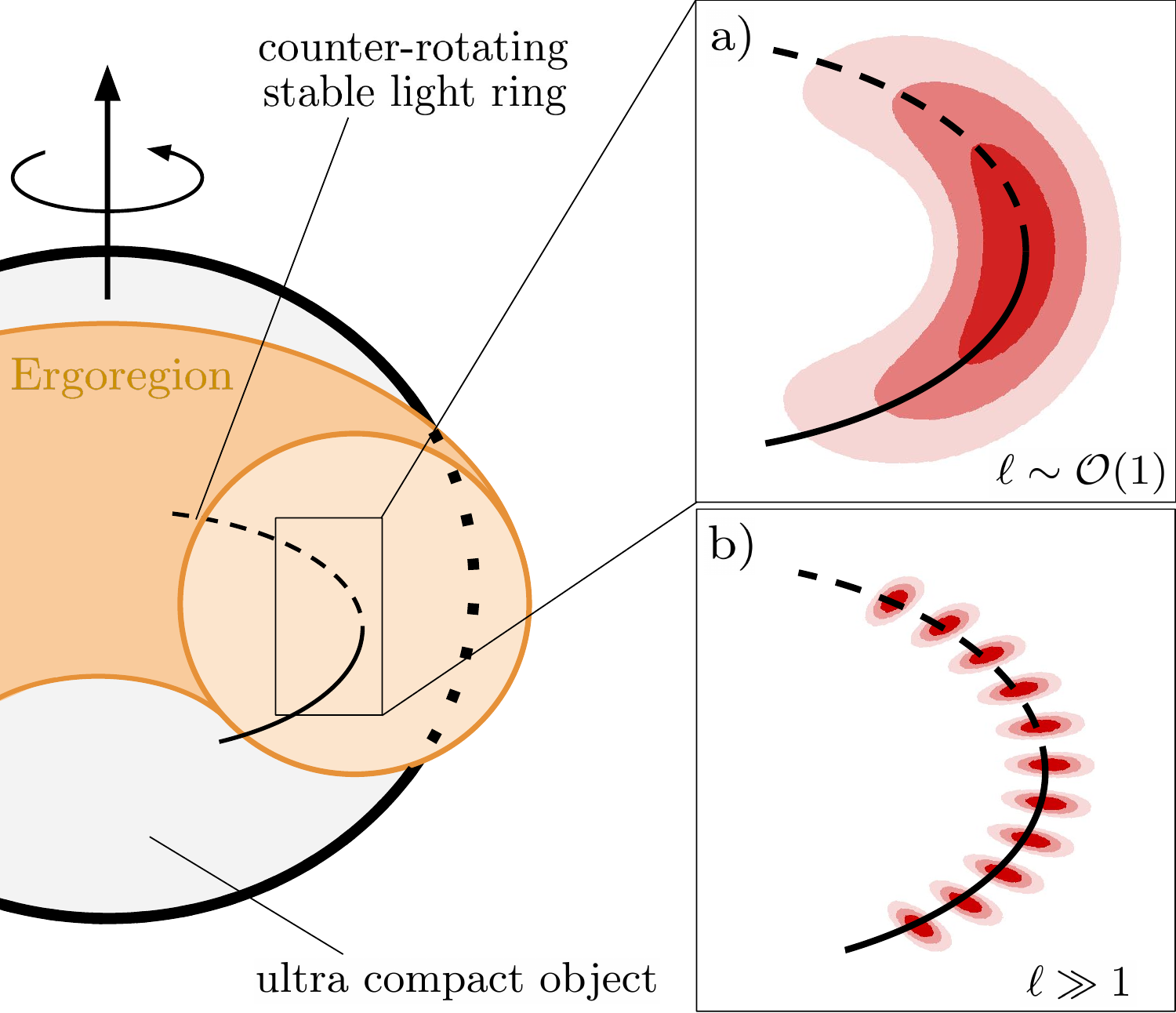}
\caption{\red{Schematic of the compact spacetime with toroidal ergoregion (a circular cross section is shown in light orange) and counter-rotating stable light ring (at the center of the ergoregion). The insets are schematically showing small-$\ell$ [panel a)] and large-$\ell$ [panel b)] linear modes. Note, the ergoregion is partially outside the object.}}
\label{fig:schematic}
\end{figure}

\noindent The background spacetime \eqref{eq:metric} exhibits an asymptotically timelike Killing vector field. This implies the existence of a conserved energy functional $E[\Psi]=E_{\rm int}+E_{\rm lin}$, up to associated outgoing boundary energy flux $\mathcal{F}=\sum_I\mathcal{F}_{\ell m}$. The energy is split into the sum of linear energies of each mode, $E_{\rm lin}=\sum_I E_{\ell m}$, and the interaction energy $E_{\rm int}$ (see~\cite{sm} for expressions). Note, in the truncated setting of eq.~\eqref{eq:KGeq} the energy $E[\Psi]$ is exactly conserved only for $\alpha=0$, while its conservation holds only up to $\mathcal{O}(\alpha^2 |\Psi|^4)$, when $\alpha\neq 0$. In principle, derivative self-interactions of the type \eqref{eq:KGeq} can lead to scalar singularities in finite time (see e.g., Ref.~\cite{fritz_1981}); however, we find this to occur only for $\alpha<0$ as discussed below. \red{Throuhout, the focus lies on $\alpha,\kappa>0$ unless stated otherwise. For later convenience, we define the frequency and growth rate of the $\ell=m=1$ mode by $\omega_R=\mathcal{F}_{11}/\mathcal{J}_{11}$ and $\omega_I=-\mathcal{F}_{11}/(2E_{11})$, respecticaly.}

We solve these coupled partial differential equations numerically in the time-domain on a finite spatial domain, imposing regularity and outgoing radiation conditions for each $\phi_{\ell m}$ at the origin and large distances, respectively. Throughout the main text, we include only modes with $\ell=m\geq 0$ up to $\ell_{\max}=23$ and dissipate energy injected into the $(\ell_{\max}+1)$-th mode. This is justified, as we find that the inclusion of the $\ell=-m>0$ or $\ell>|m|$ (for fixed $m$) modes has negligible impact on the solution as discussed in more detail in~\cite{sm}. Our focus lies on a single boson star with ergoregion and mass $M$; see ~\cite{sm}. On this background, the energy's e-folding time is $\tau^{\rm EI}_{m=1}=1.2\times 10^4M$ \cite{Siemonsen:2025wib}. Lastly, we consider two-mode small initial data: both $\phi_{11}$ and $\phi_{22}$ are set to their linear ergoregion unstable states (as constructed in Ref.~\cite{Siemonsen:2025wib}) with small amplitude ratio $A_{22}/A_{11}\leq 10^{-2}$; the amplitude is set such that the system is deep in the linear regime: $A_{11}^2\ll 1/(\kappa M^2)$ and $A_{11}^2\ll 1/\alpha$. Descriptions of the numerical methods and convergence studies can be found in \cite{sm}.

\begin{figure}
\includegraphics[width=1\linewidth]{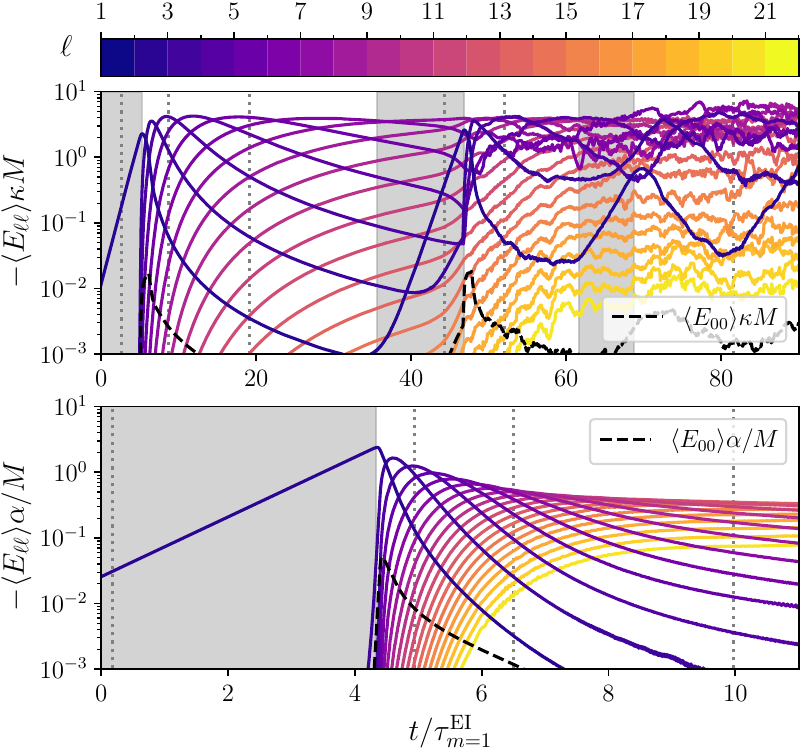}
\caption{Evolution of the energies $E_{\ell\ell}$ for $\ell\geq 0$ from the linear into the nonlinear regime. We include only potential-type self-interactions (top), or only derivative self-interactions (bottom). Here $\langle\dots\rangle$ indicates a rolling time-average. All shown energies are negative, except $E_{00}>0$ (dashed black lines). The gray shaded regions indicate the periods, when the $\ell=m=1$ mode is subject to the linear ergoregion instability. Dotted vertial lines indicate the times shown in Fig.~\ref{fig:spectrum}.}
\label{fig:mode_saturation}
\end{figure}

\textbf{Results --} \red{In Fig.~\ref{fig:schematic}, we present a schematic of the relevant structure of an ultra compact object. Specifically, the insets show qualitatively the $\phi_{\ell\ell}(t=\text{const.},r)Y_{\ell\ell}(\theta,\varphi)$ ergoregion unstable linear modes. The characteristic azimuthal length scale of the $\ell$-th mode decreases as $\ell^{-1}$, while the radial and polar extend reduced as $\ell^{-p}$, with $0<p<1$~\cite{Cardoso:2014sna}. Therefore, the larger $\ell$ the more localized the energy becomes on the stable light ring. As we detail below, we find the turbulent mechanism active in this system to transfers energy from the $\ell\sim \mathcal{O}(1)$ modes, which are driven by the ergoregion instability, towards $\ell\gg 1$, corresponding to long-lived and highly localized modes with small characteristic scales and high frequencies.}

\red{More specifically, we begin by following the linear energies} $E_{\ell\ell}$ during the evolution of the system from the linear and ergoregion instability regime, through nonlinear saturation, in Fig.~\ref{fig:mode_saturation}. In the linear phase, all modes with $\ell>2$ are sourced through the self-interactions, but with exponentially suppressed amplitudes~\footnote{Specifically, $A_{\ell\ell}\sim \kappa^{(p+q)/2} M^{p+q} A_{11}^p A_{22}^q$ and $A_{\ell\ell}\sim \alpha^{(p'+q')/2} A_{11}^{p'} A_{22}^{q'}$, where $p,q,p',q'>0$ grow roughly linearly with $\ell$.}. During the nonlinear saturation of the $\ell=m=1$ ergoregion instability around $t/\tau^{\rm EI}_{m=1}\approx 4.2$, all modes with $\ell=m > 1$ are rapidly amplified at the expense of $E_{11}$, halting the unstable process around $-\alpha E_{11}/M\sim-\kappa M E_{11}\sim\mathcal{O}(1)$. This amplification may be triggered through parametric driving of the \textit{growing} $\ell=m\geq 2$ modes by pumping energy into the \textit{decaying} $\ell=m=0$ mode analogous to findings reported in Ref.~\cite{Baryakhtar:2020gao} [we find the energy decay rate of this monopolar mode to be $\approx -4\times 10^{-3}/M\gg -2\times \text{Im}(\omega_{m=\ell\geq 1})$]. In~\cite{sm}, we discuss this pumping in more detail and explicitly demonstrate that the amplification (i) crucially hinges on mixing into the $\ell=m=0$ mode, further supporting the parametric resonance picture, and (ii) is largely independent of the initial ratio $A_{22}/A_{11}$ (assuming $A_{22}\lesssim A_{11}$), demonstrating a high degree of independence of the process from the chosen initial data. At an intuitive level, any transfer of positive energy into the monopolar mode amplifies the negative energy states satisfying the associated selection rules, i.e., the $\ell=m>1$ modes, by virtue of energy conservation.

\begin{figure}
\includegraphics[width=1\linewidth]{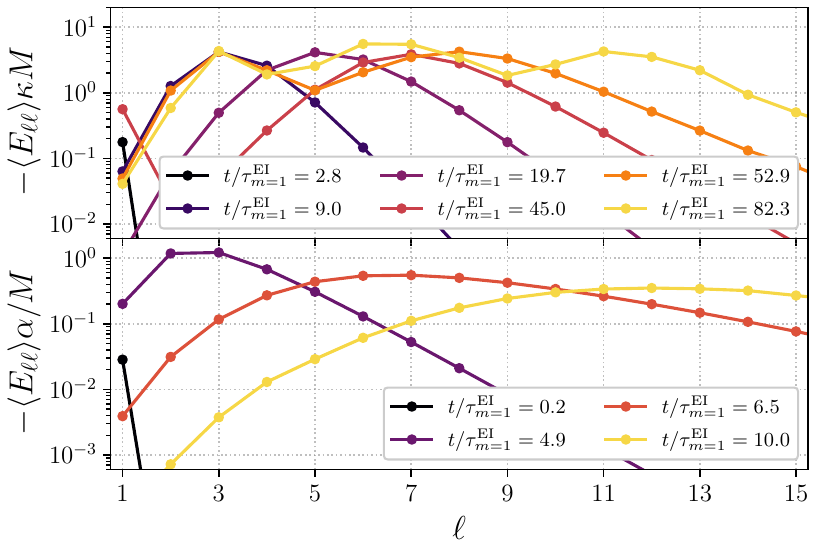}
\caption{The spectrum of linear energies $E_{\ell\ell}$ at various times during the evolution shown in Fig.~\ref{fig:mode_saturation}. The top includes only potential-type self-interactions, while the bottom only derivative self-interactions.}
\label{fig:spectrum}
\end{figure}

The subsequent evolution, $t/\tau^{\rm EI}_{m=1}\gtrsim 4$, is characterized by a weakly turbulent direct cascade from $\ell=m=1$ towards large $\ell=m>1$ modes; \red{that is, a transport of energy from modes shown in a) of Fig.~\ref{fig:schematic} towards modes shown in b).} In Fig.~\ref{fig:spectrum}, we show the magnitude of the time-averages $-\langle E_{\ell\ell}\rangle$ at a few selected times throughout the evolution. During this turbulent phase, energy is sequentially passed from one mode to the next resulting in a range of strongly coupled modes centered around a $\ell_c$ at a given time. Interestingly, modes with $\ell\lesssim\ell_c$ begin to decay as $\ell_c$ increases in time. In particular, after $-\langle E_{11}\rangle$ reached its maximum, it decays by several orders of magnitude. Focusing on the potential-type self-interactions, eventually (in this case around $t/\tau^{\rm EI}_{m=1}\approx 35$) this mode is sufficiently small and decoupled from the strongly coupled modes centered around $\ell_c$, to re-enter the linear regime and undergo exponential growth due to the ergoregion instability~\footnote{The $\ell=m=1$ mode undergoes the same instability, since the growth rates of $-\langle E_{11}\rangle$ for $t/\tau^{\rm EI}_{m=1}\lesssim 4$ and around $t/\tau^{\rm EI}_{m=1}\gtrsim 35$ are identical; see Fig.~\ref{fig:mode_saturation}.}. The saturation of this growth occurs \textit{on average} qualitatively similar to the initial saturation, generating a second sector of strongly coupled modes (see Fig.~\ref{fig:spectrum} and $t/\tau^{\rm EI}_{m=1}=52.9$). This process repeats three times in the top panel of Fig.~\ref{fig:mode_saturation}, which results in three distinct peaks in the spectrum shown in Fig.~\ref{fig:spectrum} at $t/\tau^{\rm EI}_{m=1}=82.3$. For numerical reasons (see the~\cite{sm}), we restrict the evolution with derivative self-interactions to $t/\tau^{\rm EI}_{m=1}\lesssim 10$, but expect the $\ell=m=1$ mode to exhibit qualitatively similar behavior. \red{Finally, we highlight that such a turbulent mechanism is active only across the $\ell=m$ modes; find only only exponentially suppressed mode-mixing in the polar direction at fixed $m$ (i.e., $\ell>m$) and no evidence for radial focusing at fixed $\ell$ and $m$. This implies the turbulent mechanism transfers energy only into modes most localized on the stable light ring.}

\begin{figure}
\includegraphics[width=1\linewidth]{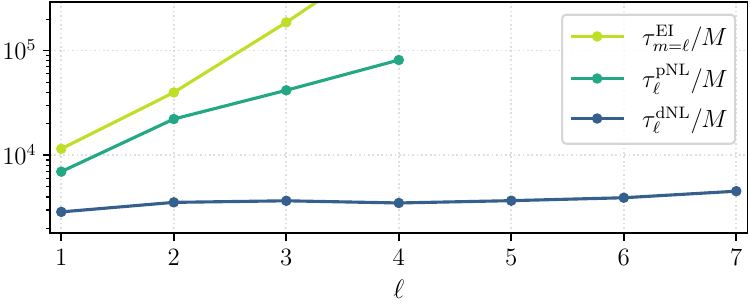}
\caption{The nonlinear energy transfer time of the potential-type self-interaction case, $\tau^{\rm pNL}_\ell$, and derivative self-interaction scenario, $\tau^{\rm dNL}_\ell$, compared to the energy e-folding timescales, $\tau^{\rm EI}_{m=\ell}$, of the linear ergoregion instability (obtained in Ref.~\cite{Siemonsen:2025wib}).}
\label{fig:nl_timescales}
\end{figure}

We quantify the energy transfer time by defining a nonlinear timescale $\tau^{\rm NL}_\ell$ as the coordinate time $t$ between the peak of $-\langle E_\ell\rangle$ and $-\langle E_{\ell+1}\rangle$ in the range separating the first and second instability saturation. These timescales are independent of the initial amplitude $A_{11}$ and practically insensitive to the ratio $A_{22}/A_{11}$. In the top panel of Fig.~\ref{fig:nl_timescales}, we compare these transfer times of both types of self-interactions to the linear e-folding growth timescales~\footnote{We restrict $\tau_\ell^{\rm pNL}$ (as defined in the caption of Fig.~\ref{fig:nl_timescales}) to modes with $\ell\leq 5$, since after the second unstable phase (around $t/\tau^{\rm EI}_{m=1}\sim 55$ in Fig.~\ref{fig:mode_saturation}), it is less clear if only immediately neighboring modes are relevant for energy transfer.}. Potential-type self-interactions lead to a slowing direct cascade~\footnote{This suggests that the turbulent process is inefficiently triggered by off-resonance energy transfer. In a system with discrete linear frequencies, such as the one considered here, the resonance condition $\omega_{\bar{m}}+\omega_{m''}=\omega_m+\omega_{m'}$ is violated leading to a slow-down of turbulent cascades (see e.g., Refs.~\cite{zakharov2012kolmogorov,nazarenko}).}. The derivative self-interactions, on the other hand, induce an efficient energy transfer with roughly constant nonlinear transfer time. \red{This may be explained as follows: derivative self-interactions have a wavenumber dependence, $\alpha|\partial\Psi|^2\sim \alpha k^2 |\Psi|^2$, which leads to qualitatively similar dynamics to potential-type self-interactions at larger scales, $\alpha k^2\sim \alpha M^{-2}\sim \kappa$, but different mode coupling at small scales $k^2\gg M^{-2}$; this likely explains the drastically different nonlinear energy transfer times.} Due to the nonlinearity, the timescales of the turbulent process, $\tau_\ell^{\rm NL}$, are inherently amplitude-dependent. That is, during the linear phase these nonlinear timescales are far larger than the e-folding timescale. As the exponential growth begins to halt, these nonlinear timescales are expected to be of the order of, or shorter than, the e-folding time of the most unstable mode's energy; this is what we observe in the top panel of Fig.~\ref{fig:nl_timescales}. \red{Once the turbulent cascade is triggered, it dominates the dynamics with timescales far shorter than the corresponding e-folding times of any of the $\ell=m>1$ modes.}

Lastly, the leading nonlinear effect prior to triggering the turbulent cascade is a perturbative shift of the unstable mode's frequency and growth rate. The deviation scales as $\omega_{R}/\text{Re}(\omega_{m=1})=1-c\beta E_{11}/M$ (where $c\in\{\alpha,\kappa\}$ and a constant $\beta >0$) in the weakly nonlinear regime (and similarly for $\omega_I$). This is in direct analogy to e.g. the black hole superradiance instability~\cite{Baryakhtar:2017ngi,May:2024npn}. Interestingly, whether these leading nonlinear effects cause an enhancement or weakening of the instability depends on the sign of $\alpha$. The choice $\alpha<0$ leads to a finite-time blow-up of the solution around $t/\tau^{\rm EI}_{m=1}\approx 5$, hence, it is not discussed further. Note, in Ref.~\cite{Siemonsen:2025wib} evidence was found that weakly nonlinear gravitational effects cause an enhancement of the ergoregion instability.

\textbf{Discussion --} In this work, we studied the nonlinear saturation of the ergoregion instability in a scalar theory with self-interactions mimicking the derivative structure of the Einstein equations \red{as a model of the gravitationally driven ergoregion instability. We found (i) that non-trivial couplings between linear modes induced by nonlinearities result in a weakly turbulent saturation of the instability, (ii) energy injected through this ergoregion unstable process at large scales (i.e., small mode numbers), cascades down towards small scales (i.e., large mode numbers) into more trapped modes primarily localized on the counter-rotating equatorial stable light ring, and (iii) the nonlinear energy transfer timescales of this turbulent process we found to be far shorter than the associated linear e-folding timescales. Physically, once the ergoregion instability reaches the nonlinear regime, the turbulent mechansim takes over, dominates the subsequent evolution of the object, and quickly funnels energy towards the stable light ring.

Large classes of spinning ultra compact and black hole mimicking objects exhibit stable light rings and are subject to the ergoregion instability. As found in this Letter, the turbulent direct cascade is ubiquitous and triggered during the nonlinear saturation of the ergoregion instability, making this a universal feature of these objects. This particularly has implications for the observational signatures of the existence of such objects~\cite{Barausse:2018vdb,Mastrogiovanni:2025ixe,Fan:2017cfw}. At small amplitudes, the gravitationally driven ergoregion instability is expected to send out nearly monochromatic gravitational waves. However, our results suggest that during the nonlinear saturation, i.e., when the gravitational wave emission is strongest, the signal may contain significant power across a large spectrum of frequencies and angular modes. This could serve as a smoking-gun signature of the ergoregion instability, and hence, may be used to probe the existence of sets of ultra compact objects motivated by high-energy and beyond the Standard Model physics~\cite{Cardoso:2019rvt,Carballo-Rubio:2025fnc,Bambi:2025wjx}.

There are various interesting aspects that deserve a more detailed study in the future. For instance, couplings between the compact object's matter may impact the turbulent direct cascade; matter itself could be subject to similar turbulent processes (e.g., fluid turbulence), but could also serve to inhibit the cascade through viscosities or other dissipation mechanisms. Furthermore, for a qualitative confirmation of the results presented here, as well as more quantitative predictions regarding the emitted gravitational wave spectrum, a complete self-consistent evolution of an ergoregion unstable compact object driven by gravitational wave emission is required. This would also help answer questions about the final state of the system, after the immediate turbulent saturation.}

\red{To \textit{speculate} about the energy spectrum of the system at late times}, the energy in each mode can be modeled by the system of equations $\dot{E}_{\ell+1,\ell+1}=E_{\ell\ell}/\tau^{\rm NL}_{\ell}-E_{\ell+1,\ell+1}/\tau^{\rm NL}_{\ell+1}$, for $\ell>1$, in a range of mode numbers up the dissipation scale $\ell_{\max}+1$. Assuming a constant-in-time energy injection rate into the $\ell=m=1$ mode (which we found to approximately hold at late times) together with stationarity of the spectrum, i.e., $\dot{E}_{\ell\ell}=0$, implies $E_{\ell\ell}\sim\tau^{\rm NL}_{\ell}$. Potential-type self-interactions lead to an increasing timescale $\tau^{\rm pNL}_\ell$ with $\ell$ (at roughly fixed energy), resulting in the energy spectrum \textit{growing} with mode number; though, this behavior is likely broken up in the eikonal limit~\footnote{In the eikonal limit $\omega_{m=\ell}\rightarrow m\omega_-$, and the resonance condition $\omega_{\bar{m}}+\omega_m=\omega_{m'}+\omega_{m''}$ is satisfying asymptotically.}. Derivative self-interactions lead to a constant transfer timescale $\tau^{\rm dNL}_\ell\sim$ const. at decreasing linear energies $-E_{\ell\ell}$, implying a decaying spectrum in $\ell$ using the above arguments. \red{ Whether, through gravitational backreaction, these perturbations collapse to black holes (and particularly at which length scale this occurs) remains open, and will depend on the precise form of $E_{\ell\ell}$. For instance, if the associated energy density scaled sufficiently steeply with $\ell$, then black hole formation could occur at small, rather than large, scales compared to the size of the compact object. We leave a more detailed treatment of the late-time spectrum and its backreaction to future work.}

Let us briefly discuss some of the shortcomings of our treatment. We justify considering only couplings between the $\ell=m\geq 0$ modes in~\cite{sm}. Beyond the treatment there, one may naively expect that the inclusion of all possible couplings either only to lead to additional, but negligible, laminar mixing, or additional pumping of energy into higher-order polar modes, which would render the state ``more'' turbulent. Lastly, these dynamics are features of the ergoregion instability and stable light ring, rather than specifics to the background spacetime chosen in this work.

\textbf{Acknowledgement --} We thank Michael Müller, Gabriele Benomio, Frans Pretorius, Will East, and Luis Lehner for many interesting discussions. We also thank the anonymous referees, who provided many useful comments. This work used \texttt{anvil} at Purdue University through allocation PHY250024 from the Advanced Cyberinfrastructure Coordination Ecosystem: Services \& Support (ACCESS) program \cite{access}, which is supported by National Science Foundation (NSF) Grants OAC-2138259, -2138286, -2138307, -2137603, and -2138296.

\bibliography{bib.bib}

\begin{thebibliography}{68}%
\makeatletter
\providecommand \@ifxundefined [1]{%
 \@ifx{#1\undefined}
}%
\providecommand \@ifnum [1]{%
 \ifnum #1\expandafter \@firstoftwo
 \else \expandafter \@secondoftwo
 \fi
}%
\providecommand \@ifx [1]{%
 \ifx #1\expandafter \@firstoftwo
 \else \expandafter \@secondoftwo
 \fi
}%
\providecommand \natexlab [1]{#1}%
\providecommand \enquote  [1]{``#1''}%
\providecommand \bibnamefont  [1]{#1}%
\providecommand \bibfnamefont [1]{#1}%
\providecommand \citenamefont [1]{#1}%
\providecommand \href@noop [0]{\@secondoftwo}%
\providecommand \href [0]{\begingroup \@sanitize@url \@href}%
\providecommand \@href[1]{\@@startlink{#1}\@@href}%
\providecommand \@@href[1]{\endgroup#1\@@endlink}%
\providecommand \@sanitize@url [0]{\catcode `\\12\catcode `\$12\catcode
  `\&12\catcode `\#12\catcode `\^12\catcode `\_12\catcode `\%12\relax}%
\providecommand \@@startlink[1]{}%
\providecommand \@@endlink[0]{}%
\providecommand \url  [0]{\begingroup\@sanitize@url \@url }%
\providecommand \@url [1]{\endgroup\@href {#1}{\urlprefix }}%
\providecommand \urlprefix  [0]{URL }%
\providecommand \Eprint [0]{\href }%
\providecommand \doibase [0]{http://dx.doi.org/}%
\providecommand \selectlanguage [0]{\@gobble}%
\providecommand \bibinfo  [0]{\@secondoftwo}%
\providecommand \bibfield  [0]{\@secondoftwo}%
\providecommand \translation [1]{[#1]}%
\providecommand \BibitemOpen [0]{}%
\providecommand \bibitemStop [0]{}%
\providecommand \bibitemNoStop [0]{.\EOS\space}%
\providecommand \EOS [0]{\spacefactor3000\relax}%
\providecommand \BibitemShut  [1]{\csname bibitem#1\endcsname}%
\let\auto@bib@innerbib\@empty
\bibitem [{\citenamefont {Cardoso}\ \emph
  {et~al.}(2014{\natexlab{a}})\citenamefont {Cardoso}, \citenamefont
  {Crispino}, \citenamefont {Macedo}, \citenamefont {Okawa},\ and\
  \citenamefont {Pani}}]{Cardoso:2014sna}%
  \BibitemOpen
  \bibfield  {author} {\bibinfo {author} {\bibfnamefont {V.}~\bibnamefont
  {Cardoso}}, \bibinfo {author} {\bibfnamefont {L.~C.~B.}\ \bibnamefont
  {Crispino}}, \bibinfo {author} {\bibfnamefont {C.~F.~B.}\ \bibnamefont
  {Macedo}}, \bibinfo {author} {\bibfnamefont {H.}~\bibnamefont {Okawa}}, \
  and\ \bibinfo {author} {\bibfnamefont {P.}~\bibnamefont {Pani}},\ }\href
  {\doibase 10.1103/PhysRevD.90.044069} {\bibfield  {journal} {\bibinfo
  {journal} {Phys. Rev. D}\ }\textbf {\bibinfo {volume} {90}},\ \bibinfo
  {pages} {044069} (\bibinfo {year} {2014}{\natexlab{a}})},\ \Eprint
  {http://arxiv.org/abs/1406.5510} {arXiv:1406.5510 [gr-qc]} \BibitemShut
  {NoStop}%
\bibitem [{\citenamefont {Friedman}(1978)}]{Friedman:1978}%
  \BibitemOpen
  \bibfield  {author} {\bibinfo {author} {\bibfnamefont {J.~L.}\ \bibnamefont
  {Friedman}},\ }\href {\doibase 10.1007/BF01196933} {\bibfield  {journal}
  {\bibinfo  {journal} {Commun. Math. Phys.}\ }\textbf {\bibinfo {volume}
  {63}},\ \bibinfo {pages} {243} (\bibinfo {year} {1978})}\BibitemShut
  {NoStop}%
\bibitem [{\citenamefont {Vilenkin}(1978)}]{Vilenkin:1978uc}%
  \BibitemOpen
  \bibfield  {author} {\bibinfo {author} {\bibfnamefont {A.}~\bibnamefont
  {Vilenkin}},\ }\href {\doibase 10.1016/0370-2693(78)90027-8} {\bibfield
  {journal} {\bibinfo  {journal} {Phys. Lett. B}\ }\textbf {\bibinfo {volume}
  {78}},\ \bibinfo {pages} {301} (\bibinfo {year} {1978})}\BibitemShut
  {NoStop}%
\bibitem [{\citenamefont {Moschidis}(2018)}]{Moschidis:2016zjy}%
  \BibitemOpen
  \bibfield  {author} {\bibinfo {author} {\bibfnamefont {G.}~\bibnamefont
  {Moschidis}},\ }\href {\doibase 10.1007/s00220-017-3010-y} {\bibfield
  {journal} {\bibinfo  {journal} {Commun. Math. Phys.}\ }\textbf {\bibinfo
  {volume} {358}},\ \bibinfo {pages} {437} (\bibinfo {year} {2018})},\ \Eprint
  {http://arxiv.org/abs/1608.02035} {arXiv:1608.02035 [math.AP]} \BibitemShut
  {NoStop}%
\bibitem [{\citenamefont {Keir}(2016)}]{Keir:2014oka}%
  \BibitemOpen
  \bibfield  {author} {\bibinfo {author} {\bibfnamefont {J.}~\bibnamefont
  {Keir}},\ }\href {\doibase 10.1088/0264-9381/33/13/135009} {\bibfield
  {journal} {\bibinfo  {journal} {Class. Quant. Grav.}\ }\textbf {\bibinfo
  {volume} {33}},\ \bibinfo {pages} {135009} (\bibinfo {year} {2016})},\
  \Eprint {http://arxiv.org/abs/1404.7036} {arXiv:1404.7036 [gr-qc]}
  \BibitemShut {NoStop}%
\bibitem [{\citenamefont {Cunha}\ \emph {et~al.}(2023)\citenamefont {Cunha},
  \citenamefont {Herdeiro}, \citenamefont {Radu},\ and\ \citenamefont
  {Sanchis-Gual}}]{Cunha:2022gde}%
  \BibitemOpen
  \bibfield  {author} {\bibinfo {author} {\bibfnamefont {P.~V.~P.}\
  \bibnamefont {Cunha}}, \bibinfo {author} {\bibfnamefont {C.}~\bibnamefont
  {Herdeiro}}, \bibinfo {author} {\bibfnamefont {E.}~\bibnamefont {Radu}}, \
  and\ \bibinfo {author} {\bibfnamefont {N.}~\bibnamefont {Sanchis-Gual}},\
  }\href {\doibase 10.1103/PhysRevLett.130.061401} {\bibfield  {journal}
  {\bibinfo  {journal} {Phys. Rev. Lett.}\ }\textbf {\bibinfo {volume} {130}},\
  \bibinfo {pages} {061401} (\bibinfo {year} {2023})},\ \Eprint
  {http://arxiv.org/abs/2207.13713} {arXiv:2207.13713 [gr-qc]} \BibitemShut
  {NoStop}%
\bibitem [{\citenamefont {Evstafyeva}\ \emph {et~al.}(2025)\citenamefont
  {Evstafyeva}, \citenamefont {Siemonsen},\ and\ \citenamefont
  {East}}]{Evstafyeva:2025mvx}%
  \BibitemOpen
  \bibfield  {author} {\bibinfo {author} {\bibfnamefont {T.}~\bibnamefont
  {Evstafyeva}}, \bibinfo {author} {\bibfnamefont {N.}~\bibnamefont
  {Siemonsen}}, \ and\ \bibinfo {author} {\bibfnamefont {W.~E.}\ \bibnamefont
  {East}},\ }\href@noop {} {\  (\bibinfo {year} {2025})},\ \Eprint
  {http://arxiv.org/abs/2508.11527} {arXiv:2508.11527 [gr-qc]} \BibitemShut
  {NoStop}%
\bibitem [{\citenamefont {Siemonsen}(2025)}]{Siemonsen:2025wib}%
  \BibitemOpen
  \bibfield  {author} {\bibinfo {author} {\bibfnamefont {N.}~\bibnamefont
  {Siemonsen}},\ }\href@noop {} {\  (\bibinfo {year} {2025})},\ \Eprint
  {http://arxiv.org/abs/2510.07468} {arXiv:2510.07468 [gr-qc]} \BibitemShut
  {NoStop}%
\bibitem [{\citenamefont {Marks}\ \emph {et~al.}(2025)\citenamefont {Marks},
  \citenamefont {Staelens}, \citenamefont {Evstafyeva},\ and\ \citenamefont
  {Sperhake}}]{Marks:2025jpt}%
  \BibitemOpen
  \bibfield  {author} {\bibinfo {author} {\bibfnamefont {G.~A.}\ \bibnamefont
  {Marks}}, \bibinfo {author} {\bibfnamefont {S.~J.}\ \bibnamefont {Staelens}},
  \bibinfo {author} {\bibfnamefont {T.}~\bibnamefont {Evstafyeva}}, \ and\
  \bibinfo {author} {\bibfnamefont {U.}~\bibnamefont {Sperhake}},\ }\href
  {\doibase 10.1103/lk48-7r2f} {\bibfield  {journal} {\bibinfo  {journal}
  {Phys. Rev. Lett.}\ }\textbf {\bibinfo {volume} {135}},\ \bibinfo {pages}
  {131402} (\bibinfo {year} {2025})},\ \Eprint
  {http://arxiv.org/abs/2504.17775} {arXiv:2504.17775 [gr-qc]} \BibitemShut
  {NoStop}%
\bibitem [{\citenamefont {Siemonsen}(2024)}]{Siemonsen:2024snb}%
  \BibitemOpen
  \bibfield  {author} {\bibinfo {author} {\bibfnamefont {N.}~\bibnamefont
  {Siemonsen}},\ }\href {\doibase 10.1103/PhysRevLett.133.031401} {\bibfield
  {journal} {\bibinfo  {journal} {Phys. Rev. Lett.}\ }\textbf {\bibinfo
  {volume} {133}},\ \bibinfo {pages} {031401} (\bibinfo {year} {2024})},\
  \Eprint {http://arxiv.org/abs/2404.14536} {arXiv:2404.14536 [gr-qc]}
  \BibitemShut {NoStop}%
\bibitem [{\citenamefont {Bizon}\ and\ \citenamefont
  {Rostworowski}(2011)}]{Bizon:2011gg}%
  \BibitemOpen
  \bibfield  {author} {\bibinfo {author} {\bibfnamefont {P.}~\bibnamefont
  {Bizon}}\ and\ \bibinfo {author} {\bibfnamefont {A.}~\bibnamefont
  {Rostworowski}},\ }\href {\doibase 10.1103/PhysRevLett.107.031102} {\bibfield
   {journal} {\bibinfo  {journal} {Phys. Rev. Lett.}\ }\textbf {\bibinfo
  {volume} {107}},\ \bibinfo {pages} {031102} (\bibinfo {year} {2011})},\
  \Eprint {http://arxiv.org/abs/1104.3702} {arXiv:1104.3702 [gr-qc]}
  \BibitemShut {NoStop}%
\bibitem [{\citenamefont {Jalmuzna}\ \emph {et~al.}(2011)\citenamefont
  {Jalmuzna}, \citenamefont {Rostworowski},\ and\ \citenamefont
  {Bizon}}]{Jalmuzna:2011qw}%
  \BibitemOpen
  \bibfield  {author} {\bibinfo {author} {\bibfnamefont {J.}~\bibnamefont
  {Jalmuzna}}, \bibinfo {author} {\bibfnamefont {A.}~\bibnamefont
  {Rostworowski}}, \ and\ \bibinfo {author} {\bibfnamefont {P.}~\bibnamefont
  {Bizon}},\ }\href {\doibase 10.1103/PhysRevD.84.085021} {\bibfield  {journal}
  {\bibinfo  {journal} {Phys. Rev. D}\ }\textbf {\bibinfo {volume} {84}},\
  \bibinfo {pages} {085021} (\bibinfo {year} {2011})},\ \Eprint
  {http://arxiv.org/abs/1108.4539} {arXiv:1108.4539 [gr-qc]} \BibitemShut
  {NoStop}%
\bibitem [{\citenamefont {Buchel}\ \emph {et~al.}(2012)\citenamefont {Buchel},
  \citenamefont {Lehner},\ and\ \citenamefont {Liebling}}]{Buchel:2012uh}%
  \BibitemOpen
  \bibfield  {author} {\bibinfo {author} {\bibfnamefont {A.}~\bibnamefont
  {Buchel}}, \bibinfo {author} {\bibfnamefont {L.}~\bibnamefont {Lehner}}, \
  and\ \bibinfo {author} {\bibfnamefont {S.~L.}\ \bibnamefont {Liebling}},\
  }\href {\doibase 10.1103/PhysRevD.86.123011} {\bibfield  {journal} {\bibinfo
  {journal} {Phys. Rev. D}\ }\textbf {\bibinfo {volume} {86}},\ \bibinfo
  {pages} {123011} (\bibinfo {year} {2012})},\ \Eprint
  {http://arxiv.org/abs/1210.0890} {arXiv:1210.0890 [gr-qc]} \BibitemShut
  {NoStop}%
\bibitem [{\citenamefont {Balasubramanian}\ \emph {et~al.}(2014)\citenamefont
  {Balasubramanian}, \citenamefont {Buchel}, \citenamefont {Green},
  \citenamefont {Lehner},\ and\ \citenamefont
  {Liebling}}]{Balasubramanian:2014cja}%
  \BibitemOpen
  \bibfield  {author} {\bibinfo {author} {\bibfnamefont {V.}~\bibnamefont
  {Balasubramanian}}, \bibinfo {author} {\bibfnamefont {A.}~\bibnamefont
  {Buchel}}, \bibinfo {author} {\bibfnamefont {S.~R.}\ \bibnamefont {Green}},
  \bibinfo {author} {\bibfnamefont {L.}~\bibnamefont {Lehner}}, \ and\ \bibinfo
  {author} {\bibfnamefont {S.~L.}\ \bibnamefont {Liebling}},\ }\href {\doibase
  10.1103/PhysRevLett.113.071601} {\bibfield  {journal} {\bibinfo  {journal}
  {Phys. Rev. Lett.}\ }\textbf {\bibinfo {volume} {113}},\ \bibinfo {pages}
  {071601} (\bibinfo {year} {2014})},\ \Eprint {http://arxiv.org/abs/1403.6471}
  {arXiv:1403.6471 [hep-th]} \BibitemShut {NoStop}%
\bibitem [{\citenamefont {Craps}\ \emph {et~al.}(2014)\citenamefont {Craps},
  \citenamefont {Evnin},\ and\ \citenamefont {Vanhoof}}]{Craps:2014vaa}%
  \BibitemOpen
  \bibfield  {author} {\bibinfo {author} {\bibfnamefont {B.}~\bibnamefont
  {Craps}}, \bibinfo {author} {\bibfnamefont {O.}~\bibnamefont {Evnin}}, \ and\
  \bibinfo {author} {\bibfnamefont {J.}~\bibnamefont {Vanhoof}},\ }\href
  {\doibase 10.1007/JHEP10(2014)048} {\bibfield  {journal} {\bibinfo  {journal}
  {JHEP}\ }\textbf {\bibinfo {volume} {10}},\ \bibinfo {pages} {048} (\bibinfo
  {year} {2014})},\ \Eprint {http://arxiv.org/abs/1407.6273} {arXiv:1407.6273
  [gr-qc]} \BibitemShut {NoStop}%
\bibitem [{\citenamefont {Bizo{\'n}}\ \emph {et~al.}(2015)\citenamefont
  {Bizo{\'n}}, \citenamefont {Maliborski},\ and\ \citenamefont
  {Rostworowski}}]{Bizon:2015pfa}%
  \BibitemOpen
  \bibfield  {author} {\bibinfo {author} {\bibfnamefont {P.}~\bibnamefont
  {Bizo{\'n}}}, \bibinfo {author} {\bibfnamefont {M.}~\bibnamefont
  {Maliborski}}, \ and\ \bibinfo {author} {\bibfnamefont {A.}~\bibnamefont
  {Rostworowski}},\ }\href {\doibase 10.1103/PhysRevLett.115.081103} {\bibfield
   {journal} {\bibinfo  {journal} {Phys. Rev. Lett.}\ }\textbf {\bibinfo
  {volume} {115}},\ \bibinfo {pages} {081103} (\bibinfo {year} {2015})},\
  \Eprint {http://arxiv.org/abs/1506.03519} {arXiv:1506.03519 [gr-qc]}
  \BibitemShut {NoStop}%
\bibitem [{\citenamefont {Evnin}(2021)}]{Evnin:2021buq}%
  \BibitemOpen
  \bibfield  {author} {\bibinfo {author} {\bibfnamefont {O.}~\bibnamefont
  {Evnin}},\ }\href {\doibase 10.1088/1361-6382/ac1b46} {\bibfield  {journal}
  {\bibinfo  {journal} {Class. Quant. Grav.}\ }\textbf {\bibinfo {volume}
  {38}},\ \bibinfo {pages} {203001} (\bibinfo {year} {2021})},\ \Eprint
  {http://arxiv.org/abs/2104.09797} {arXiv:2104.09797 [gr-qc]} \BibitemShut
  {NoStop}%
\bibitem [{\citenamefont {Galtier}\ and\ \citenamefont
  {Nazarenko}(2017)}]{Galtier:2017mve}%
  \BibitemOpen
  \bibfield  {author} {\bibinfo {author} {\bibfnamefont {S.}~\bibnamefont
  {Galtier}}\ and\ \bibinfo {author} {\bibfnamefont {S.~V.}\ \bibnamefont
  {Nazarenko}},\ }\href {\doibase 10.1103/PhysRevLett.119.221101} {\bibfield
  {journal} {\bibinfo  {journal} {Phys. Rev. Lett.}\ }\textbf {\bibinfo
  {volume} {119}},\ \bibinfo {pages} {221101} (\bibinfo {year} {2017})},\
  \Eprint {http://arxiv.org/abs/1703.09069} {arXiv:1703.09069 [gr-qc]}
  \BibitemShut {NoStop}%
\bibitem [{\citenamefont {Galtier}\ and\ \citenamefont
  {Nazarenko}(2021)}]{Galtier:2021ovg}%
  \BibitemOpen
  \bibfield  {author} {\bibinfo {author} {\bibfnamefont {S.}~\bibnamefont
  {Galtier}}\ and\ \bibinfo {author} {\bibfnamefont {S.~V.}\ \bibnamefont
  {Nazarenko}},\ }\href {\doibase 10.1103/PhysRevLett.127.131101} {\bibfield
  {journal} {\bibinfo  {journal} {Phys. Rev. Lett.}\ }\textbf {\bibinfo
  {volume} {127}},\ \bibinfo {pages} {131101} (\bibinfo {year} {2021})},\
  \Eprint {http://arxiv.org/abs/2108.09158} {arXiv:2108.09158 [gr-qc]}
  \BibitemShut {NoStop}%
\bibitem [{\citenamefont {Benomio}(2021)}]{Benomio:2018ivy}%
  \BibitemOpen
  \bibfield  {author} {\bibinfo {author} {\bibfnamefont {G.}~\bibnamefont
  {Benomio}},\ }\href {\doibase 10.2140/apde.2021.14.2427} {\bibfield
  {journal} {\bibinfo  {journal} {Anal. Part. Diff. Eq.}\ }\textbf {\bibinfo
  {volume} {14}},\ \bibinfo {pages} {2427} (\bibinfo {year} {2021})},\ \Eprint
  {http://arxiv.org/abs/1809.07795} {arXiv:1809.07795 [gr-qc]} \BibitemShut
  {NoStop}%
\bibitem [{\citenamefont {Bhattacharyya}\ \emph {et~al.}(2008)\citenamefont
  {Bhattacharyya}, \citenamefont {Hubeny}, \citenamefont {Minwalla},\ and\
  \citenamefont {Rangamani}}]{Bhattacharyya:2007vjd}%
  \BibitemOpen
  \bibfield  {author} {\bibinfo {author} {\bibfnamefont {S.}~\bibnamefont
  {Bhattacharyya}}, \bibinfo {author} {\bibfnamefont {V.~E.}\ \bibnamefont
  {Hubeny}}, \bibinfo {author} {\bibfnamefont {S.}~\bibnamefont {Minwalla}}, \
  and\ \bibinfo {author} {\bibfnamefont {M.}~\bibnamefont {Rangamani}},\ }\href
  {\doibase 10.1088/1126-6708/2008/02/045} {\bibfield  {journal} {\bibinfo
  {journal} {JHEP}\ }\textbf {\bibinfo {volume} {02}},\ \bibinfo {pages} {045}
  (\bibinfo {year} {2008})},\ \Eprint {http://arxiv.org/abs/0712.2456}
  {arXiv:0712.2456 [hep-th]} \BibitemShut {NoStop}%
\bibitem [{\citenamefont {Carrasco}\ \emph {et~al.}(2012)\citenamefont
  {Carrasco}, \citenamefont {Lehner}, \citenamefont {Myers}, \citenamefont
  {Reula},\ and\ \citenamefont {Singh}}]{Carrasco:2012nf}%
  \BibitemOpen
  \bibfield  {author} {\bibinfo {author} {\bibfnamefont {F.}~\bibnamefont
  {Carrasco}}, \bibinfo {author} {\bibfnamefont {L.}~\bibnamefont {Lehner}},
  \bibinfo {author} {\bibfnamefont {R.~C.}\ \bibnamefont {Myers}}, \bibinfo
  {author} {\bibfnamefont {O.}~\bibnamefont {Reula}}, \ and\ \bibinfo {author}
  {\bibfnamefont {A.}~\bibnamefont {Singh}},\ }\href {\doibase
  10.1103/PhysRevD.86.126006} {\bibfield  {journal} {\bibinfo  {journal} {Phys.
  Rev. D}\ }\textbf {\bibinfo {volume} {86}},\ \bibinfo {pages} {126006}
  (\bibinfo {year} {2012})},\ \Eprint {http://arxiv.org/abs/1210.6702}
  {arXiv:1210.6702 [hep-th]} \BibitemShut {NoStop}%
\bibitem [{\citenamefont {Adams}\ \emph {et~al.}(2014)\citenamefont {Adams},
  \citenamefont {Chesler},\ and\ \citenamefont {Liu}}]{Adams:2013vsa}%
  \BibitemOpen
  \bibfield  {author} {\bibinfo {author} {\bibfnamefont {A.}~\bibnamefont
  {Adams}}, \bibinfo {author} {\bibfnamefont {P.~M.}\ \bibnamefont {Chesler}},
  \ and\ \bibinfo {author} {\bibfnamefont {H.}~\bibnamefont {Liu}},\ }\href
  {\doibase 10.1103/PhysRevLett.112.151602} {\bibfield  {journal} {\bibinfo
  {journal} {Phys. Rev. Lett.}\ }\textbf {\bibinfo {volume} {112}},\ \bibinfo
  {pages} {151602} (\bibinfo {year} {2014})},\ \Eprint
  {http://arxiv.org/abs/1307.7267} {arXiv:1307.7267 [hep-th]} \BibitemShut
  {NoStop}%
\bibitem [{\citenamefont {Green}\ \emph {et~al.}(2014)\citenamefont {Green},
  \citenamefont {Carrasco},\ and\ \citenamefont {Lehner}}]{Green:2013zba}%
  \BibitemOpen
  \bibfield  {author} {\bibinfo {author} {\bibfnamefont {S.~R.}\ \bibnamefont
  {Green}}, \bibinfo {author} {\bibfnamefont {F.}~\bibnamefont {Carrasco}}, \
  and\ \bibinfo {author} {\bibfnamefont {L.}~\bibnamefont {Lehner}},\ }\href
  {\doibase 10.1103/PhysRevX.4.011001} {\bibfield  {journal} {\bibinfo
  {journal} {Phys. Rev. X}\ }\textbf {\bibinfo {volume} {4}},\ \bibinfo {pages}
  {011001} (\bibinfo {year} {2014})},\ \Eprint {http://arxiv.org/abs/1309.7940}
  {arXiv:1309.7940 [hep-th]} \BibitemShut {NoStop}%
\bibitem [{\citenamefont {Krynicki}\ \emph {et~al.}(2025)\citenamefont
  {Krynicki}, \citenamefont {Wu},\ and\ \citenamefont
  {Most}}]{Krynicki:2025fzi}%
  \BibitemOpen
  \bibfield  {author} {\bibinfo {author} {\bibfnamefont {H.}~\bibnamefont
  {Krynicki}}, \bibinfo {author} {\bibfnamefont {J.}~\bibnamefont {Wu}}, \ and\
  \bibinfo {author} {\bibfnamefont {E.~R.}\ \bibnamefont {Most}},\ }\href@noop
  {} {\  (\bibinfo {year} {2025})},\ \Eprint {http://arxiv.org/abs/2509.19769}
  {arXiv:2509.19769 [gr-qc]} \BibitemShut {NoStop}%
\bibitem [{\citenamefont {Gay}\ and\ \citenamefont
  {Galtier}(2024)}]{Gay:2024kay}%
  \BibitemOpen
  \bibfield  {author} {\bibinfo {author} {\bibfnamefont {B.}~\bibnamefont
  {Gay}}\ and\ \bibinfo {author} {\bibfnamefont {S.}~\bibnamefont {Galtier}},\
  }\href {\doibase 10.1103/PhysRevD.109.083531} {\bibfield  {journal} {\bibinfo
   {journal} {Phys. Rev. D}\ }\textbf {\bibinfo {volume} {109}},\ \bibinfo
  {pages} {083531} (\bibinfo {year} {2024})},\ \Eprint
  {http://arxiv.org/abs/2402.05614} {arXiv:2402.05614 [gr-qc]} \BibitemShut
  {NoStop}%
\bibitem [{\citenamefont {Liang}\ \emph {et~al.}(2025)\citenamefont {Liang},
  \citenamefont {Xu}, \citenamefont {Du}, \citenamefont {Cheng}, \citenamefont
  {Wang},\ and\ \citenamefont {Luo}}]{Liang:2025lek}%
  \BibitemOpen
  \bibfield  {author} {\bibinfo {author} {\bibfnamefont {J.}~\bibnamefont
  {Liang}}, \bibinfo {author} {\bibfnamefont {P.}~\bibnamefont {Xu}}, \bibinfo
  {author} {\bibfnamefont {M.}~\bibnamefont {Du}}, \bibinfo {author}
  {\bibfnamefont {Y.}~\bibnamefont {Cheng}}, \bibinfo {author} {\bibfnamefont
  {Z.}~\bibnamefont {Wang}}, \ and\ \bibinfo {author} {\bibfnamefont
  {Z.}~\bibnamefont {Luo}},\ }\href@noop {} {\  (\bibinfo {year} {2025})},\
  \Eprint {http://arxiv.org/abs/2510.03711} {arXiv:2510.03711 [gr-qc]}
  \BibitemShut {NoStop}%
\bibitem [{\citenamefont {Andrade}\ \emph {et~al.}(2021)\citenamefont
  {Andrade}, \citenamefont {Pantelidou}, \citenamefont {Sonner},\ and\
  \citenamefont {Withers}}]{Andrade:2019rpn}%
  \BibitemOpen
  \bibfield  {author} {\bibinfo {author} {\bibfnamefont {T.}~\bibnamefont
  {Andrade}}, \bibinfo {author} {\bibfnamefont {C.}~\bibnamefont {Pantelidou}},
  \bibinfo {author} {\bibfnamefont {J.}~\bibnamefont {Sonner}}, \ and\ \bibinfo
  {author} {\bibfnamefont {B.}~\bibnamefont {Withers}},\ }\href {\doibase
  10.1007/JHEP07(2021)063} {\bibfield  {journal} {\bibinfo  {journal} {JHEP}\
  }\textbf {\bibinfo {volume} {07}},\ \bibinfo {pages} {063} (\bibinfo {year}
  {2021})},\ \Eprint {http://arxiv.org/abs/1912.00032} {arXiv:1912.00032
  [hep-th]} \BibitemShut {NoStop}%
\bibitem [{\citenamefont {Yang}\ \emph {et~al.}(2015)\citenamefont {Yang},
  \citenamefont {Zimmerman},\ and\ \citenamefont {Lehner}}]{Yang:2014tla}%
  \BibitemOpen
  \bibfield  {author} {\bibinfo {author} {\bibfnamefont {H.}~\bibnamefont
  {Yang}}, \bibinfo {author} {\bibfnamefont {A.}~\bibnamefont {Zimmerman}}, \
  and\ \bibinfo {author} {\bibfnamefont {L.}~\bibnamefont {Lehner}},\ }\href
  {\doibase 10.1103/PhysRevLett.114.081101} {\bibfield  {journal} {\bibinfo
  {journal} {Phys. Rev. Lett.}\ }\textbf {\bibinfo {volume} {114}},\ \bibinfo
  {pages} {081101} (\bibinfo {year} {2015})},\ \Eprint
  {http://arxiv.org/abs/1402.4859} {arXiv:1402.4859 [gr-qc]} \BibitemShut
  {NoStop}%
\bibitem [{\citenamefont {Iuliano}\ \emph {et~al.}(2025)\citenamefont
  {Iuliano}, \citenamefont {Hollands}, \citenamefont {Green},\ and\
  \citenamefont {Zimmerman}}]{Iuliano:2024ogr}%
  \BibitemOpen
  \bibfield  {author} {\bibinfo {author} {\bibfnamefont {C.}~\bibnamefont
  {Iuliano}}, \bibinfo {author} {\bibfnamefont {S.}~\bibnamefont {Hollands}},
  \bibinfo {author} {\bibfnamefont {S.~R.}\ \bibnamefont {Green}}, \ and\
  \bibinfo {author} {\bibfnamefont {P.}~\bibnamefont {Zimmerman}},\ }\href
  {\doibase 10.1103/PhysRevD.111.124038} {\bibfield  {journal} {\bibinfo
  {journal} {Phys. Rev. D}\ }\textbf {\bibinfo {volume} {111}},\ \bibinfo
  {pages} {124038} (\bibinfo {year} {2025})},\ \Eprint
  {http://arxiv.org/abs/2412.02821} {arXiv:2412.02821 [gr-qc]} \BibitemShut
  {NoStop}%
\bibitem [{\citenamefont {Figueras}\ and\ \citenamefont
  {Rossi}(2025)}]{Figueras:2023ihz}%
  \BibitemOpen
  \bibfield  {author} {\bibinfo {author} {\bibfnamefont {P.}~\bibnamefont
  {Figueras}}\ and\ \bibinfo {author} {\bibfnamefont {L.}~\bibnamefont
  {Rossi}},\ }\href {\doibase 10.1007/JHEP06(2025)107} {\bibfield  {journal}
  {\bibinfo  {journal} {JHEP}\ }\textbf {\bibinfo {volume} {06}},\ \bibinfo
  {pages} {107} (\bibinfo {year} {2025})},\ \Eprint
  {http://arxiv.org/abs/2311.14167} {arXiv:2311.14167 [hep-th]} \BibitemShut
  {NoStop}%
\bibitem [{\citenamefont {Ma}\ \emph {et~al.}(2025)\citenamefont {Ma},
  \citenamefont {Lehner}, \citenamefont {Yang}, \citenamefont {Kidder},
  \citenamefont {Pfeiffer},\ and\ \citenamefont {Scheel}}]{Ma:2025rnv}%
  \BibitemOpen
  \bibfield  {author} {\bibinfo {author} {\bibfnamefont {S.}~\bibnamefont
  {Ma}}, \bibinfo {author} {\bibfnamefont {L.}~\bibnamefont {Lehner}}, \bibinfo
  {author} {\bibfnamefont {H.}~\bibnamefont {Yang}}, \bibinfo {author}
  {\bibfnamefont {L.~E.}\ \bibnamefont {Kidder}}, \bibinfo {author}
  {\bibfnamefont {H.~P.}\ \bibnamefont {Pfeiffer}}, \ and\ \bibinfo {author}
  {\bibfnamefont {M.~A.}\ \bibnamefont {Scheel}},\ }\href@noop {} {\  (\bibinfo
  {year} {2025})},\ \Eprint {http://arxiv.org/abs/2508.13294} {arXiv:2508.13294
  [gr-qc]} \BibitemShut {NoStop}%
\bibitem [{\citenamefont {Detweiler}(1977)}]{Detweiler:1977gy}%
  \BibitemOpen
  \bibfield  {author} {\bibinfo {author} {\bibfnamefont {S.~L.}\ \bibnamefont
  {Detweiler}},\ }\href {\doibase 10.1098/rspa.1977.0005} {\bibfield  {journal}
  {\bibinfo  {journal} {Proc. Roy. Soc. Lond. A}\ }\textbf {\bibinfo {volume}
  {352}},\ \bibinfo {pages} {381} (\bibinfo {year} {1977})}\BibitemShut
  {NoStop}%
\bibitem [{\citenamefont {Hod}(2008)}]{Hod:2008zz}%
  \BibitemOpen
  \bibfield  {author} {\bibinfo {author} {\bibfnamefont {S.}~\bibnamefont
  {Hod}},\ }\href {\doibase 10.1103/PhysRevD.78.084035} {\bibfield  {journal}
  {\bibinfo  {journal} {Phys. Rev. D}\ }\textbf {\bibinfo {volume} {78}},\
  \bibinfo {pages} {084035} (\bibinfo {year} {2008})},\ \Eprint
  {http://arxiv.org/abs/0811.3806} {arXiv:0811.3806 [gr-qc]} \BibitemShut
  {NoStop}%
\bibitem [{\citenamefont {Yang}\ \emph {et~al.}(2013)\citenamefont {Yang},
  \citenamefont {Zhang}, \citenamefont {Zimmerman}, \citenamefont {Nichols},
  \citenamefont {Berti},\ and\ \citenamefont {Chen}}]{Yang:2012pj}%
  \BibitemOpen
  \bibfield  {author} {\bibinfo {author} {\bibfnamefont {H.}~\bibnamefont
  {Yang}}, \bibinfo {author} {\bibfnamefont {F.}~\bibnamefont {Zhang}},
  \bibinfo {author} {\bibfnamefont {A.}~\bibnamefont {Zimmerman}}, \bibinfo
  {author} {\bibfnamefont {D.~A.}\ \bibnamefont {Nichols}}, \bibinfo {author}
  {\bibfnamefont {E.}~\bibnamefont {Berti}}, \ and\ \bibinfo {author}
  {\bibfnamefont {Y.}~\bibnamefont {Chen}},\ }\href {\doibase
  10.1103/PhysRevD.87.041502} {\bibfield  {journal} {\bibinfo  {journal} {Phys.
  Rev. D}\ }\textbf {\bibinfo {volume} {87}},\ \bibinfo {pages} {041502}
  (\bibinfo {year} {2013})},\ \Eprint {http://arxiv.org/abs/1212.3271}
  {arXiv:1212.3271 [gr-qc]} \BibitemShut {NoStop}%
\bibitem [{\citenamefont {Cardoso}\ \emph
  {et~al.}(2014{\natexlab{b}})\citenamefont {Cardoso}, \citenamefont {Dias},
  \citenamefont {Hartnett}, \citenamefont {Lehner},\ and\ \citenamefont
  {Santos}}]{Cardoso:2013pza}%
  \BibitemOpen
  \bibfield  {author} {\bibinfo {author} {\bibfnamefont {V.}~\bibnamefont
  {Cardoso}}, \bibinfo {author} {\bibfnamefont {{\'O}.~J.~C.}\ \bibnamefont
  {Dias}}, \bibinfo {author} {\bibfnamefont {G.~S.}\ \bibnamefont {Hartnett}},
  \bibinfo {author} {\bibfnamefont {L.}~\bibnamefont {Lehner}}, \ and\ \bibinfo
  {author} {\bibfnamefont {J.~E.}\ \bibnamefont {Santos}},\ }\href {\doibase
  10.1007/JHEP04(2014)183} {\bibfield  {journal} {\bibinfo  {journal} {JHEP}\
  }\textbf {\bibinfo {volume} {04}},\ \bibinfo {pages} {183} (\bibinfo {year}
  {2014}{\natexlab{b}})},\ \Eprint {http://arxiv.org/abs/1312.5323}
  {arXiv:1312.5323 [hep-th]} \BibitemShut {NoStop}%
\bibitem [{\citenamefont {Holzegel}\ and\ \citenamefont
  {Smulevici}(2014)}]{Holzegel:2013kna}%
  \BibitemOpen
  \bibfield  {author} {\bibinfo {author} {\bibfnamefont {G.}~\bibnamefont
  {Holzegel}}\ and\ \bibinfo {author} {\bibfnamefont {J.}~\bibnamefont
  {Smulevici}},\ }\href {\doibase 10.2140/apde.2014.7.1057} {\bibfield
  {journal} {\bibinfo  {journal} {Anal. Part. Diff. Eq.}\ }\textbf {\bibinfo
  {volume} {7}},\ \bibinfo {pages} {1057} (\bibinfo {year} {2014})},\ \Eprint
  {http://arxiv.org/abs/1303.5944} {arXiv:1303.5944 [gr-qc]} \BibitemShut
  {NoStop}%
\bibitem [{\citenamefont {Benomio}\ \emph {et~al.}(2024)\citenamefont
  {Benomio}, \citenamefont {C\'ardenas-Avenda\~no}, \citenamefont {Pretorius},\
  and\ \citenamefont {Sullivan}}]{Benomio:2024lev}%
  \BibitemOpen
  \bibfield  {author} {\bibinfo {author} {\bibfnamefont {G.}~\bibnamefont
  {Benomio}}, \bibinfo {author} {\bibfnamefont {A.}~\bibnamefont
  {C\'ardenas-Avenda\~no}}, \bibinfo {author} {\bibfnamefont {F.}~\bibnamefont
  {Pretorius}}, \ and\ \bibinfo {author} {\bibfnamefont {A.}~\bibnamefont
  {Sullivan}},\ }\href@noop {} {\  (\bibinfo {year} {2024})},\ \Eprint
  {http://arxiv.org/abs/2411.17445} {arXiv:2411.17445 [gr-qc]} \BibitemShut
  {NoStop}%
\bibitem [{\citenamefont {Redondo-Yuste}\ and\ \citenamefont
  {C\'ardenas-Avenda\~no}(2025)}]{Redondo-Yuste:2025hlv}%
  \BibitemOpen
  \bibfield  {author} {\bibinfo {author} {\bibfnamefont {J.}~\bibnamefont
  {Redondo-Yuste}}\ and\ \bibinfo {author} {\bibfnamefont {A.}~\bibnamefont
  {C\'ardenas-Avenda\~no}},\ }\href@noop {} {\  (\bibinfo {year} {2025})},\
  \Eprint {http://arxiv.org/abs/2502.18643} {arXiv:2502.18643 [gr-qc]}
  \BibitemShut {NoStop}%
\bibitem [{\citenamefont {Zakharov}\ \emph {et~al.}(2012)\citenamefont
  {Zakharov}, \citenamefont {L'vov},\ and\ \citenamefont
  {Falkovich}}]{zakharov2012kolmogorov}%
  \BibitemOpen
  \bibfield  {author} {\bibinfo {author} {\bibfnamefont {V.}~\bibnamefont
  {Zakharov}}, \bibinfo {author} {\bibfnamefont {V.}~\bibnamefont {L'vov}}, \
  and\ \bibinfo {author} {\bibfnamefont {G.}~\bibnamefont {Falkovich}},\ }\href
  {https://link.springer.com/book/10.1007/978-3-642-50052-7} {\emph {\bibinfo
  {title} {Kolmogorov Spectra of Turbulence I: Wave Turbulence}}},\ Springer
  Series in Nonlinear Dynamics\ (\bibinfo  {publisher} {Springer Berlin
  Heidelberg},\ \bibinfo {year} {2012})\BibitemShut {NoStop}%
\bibitem [{\citenamefont {Galtier}(2022)}]{Galtier_2022}%
  \BibitemOpen
  \bibfield  {author} {\bibinfo {author} {\bibfnamefont {S.}~\bibnamefont
  {Galtier}},\ }\href@noop {} {\emph {\bibinfo {title} {Physics of Wave
  Turbulence}}}\ (\bibinfo  {publisher} {Cambridge University Press},\ \bibinfo
  {year} {2022})\BibitemShut {NoStop}%
\bibitem [{\citenamefont {Nazarenko}(2011)}]{nazarenko}%
  \BibitemOpen
  \bibfield  {author} {\bibinfo {author} {\bibfnamefont {S.}~\bibnamefont
  {Nazarenko}},\ }\href
  {https://link.springer.com/book/10.1007/978-3-642-15942-8} {\emph {\bibinfo
  {title} {Wave Turbulence}}},\ Lecture Notes in Physics\ (\bibinfo
  {publisher} {Springer Berlin Heidelberg},\ \bibinfo {year}
  {2011})\BibitemShut {NoStop}%
\bibitem [{\citenamefont {Cardoso}\ and\ \citenamefont
  {Pani}(2019)}]{Cardoso:2019rvt}%
  \BibitemOpen
  \bibfield  {author} {\bibinfo {author} {\bibfnamefont {V.}~\bibnamefont
  {Cardoso}}\ and\ \bibinfo {author} {\bibfnamefont {P.}~\bibnamefont {Pani}},\
  }\href {\doibase 10.1007/s41114-019-0020-4} {\bibfield  {journal} {\bibinfo
  {journal} {Living Rev. Rel.}\ }\textbf {\bibinfo {volume} {22}},\ \bibinfo
  {pages} {4} (\bibinfo {year} {2019})},\ \Eprint
  {http://arxiv.org/abs/1904.05363} {arXiv:1904.05363 [gr-qc]} \BibitemShut
  {NoStop}%
\bibitem [{\citenamefont {Bambi}\ \emph {et~al.}(2025)\citenamefont {Bambi}
  \emph {et~al.}}]{Bambi:2025wjx}%
  \BibitemOpen
  \bibfield  {author} {\bibinfo {author} {\bibfnamefont {C.}~\bibnamefont
  {Bambi}} \emph {et~al.}\ }(\bibinfo {year} {2025})\ \Eprint
  {http://arxiv.org/abs/2505.09014} {arXiv:2505.09014 [gr-qc]} \BibitemShut
  {NoStop}%
\bibitem [{\citenamefont {Carballo-Rubio}\ \emph {et~al.}(2025)\citenamefont
  {Carballo-Rubio} \emph {et~al.}}]{Carballo-Rubio:2025fnc}%
  \BibitemOpen
  \bibfield  {author} {\bibinfo {author} {\bibfnamefont {R.}~\bibnamefont
  {Carballo-Rubio}} \emph {et~al.},\ }\href {\doibase
  10.1088/1475-7516/2025/05/003} {\bibfield  {journal} {\bibinfo  {journal}
  {JCAP}\ }\textbf {\bibinfo {volume} {05}},\ \bibinfo {pages} {003} (\bibinfo
  {year} {2025})},\ \Eprint {http://arxiv.org/abs/2501.05505} {arXiv:2501.05505
  [gr-qc]} \BibitemShut {NoStop}%
\bibitem [{\citenamefont {Barausse}\ \emph {et~al.}(2018)\citenamefont
  {Barausse}, \citenamefont {Brito}, \citenamefont {Cardoso}, \citenamefont
  {Dvorkin},\ and\ \citenamefont {Pani}}]{Barausse:2018vdb}%
  \BibitemOpen
  \bibfield  {author} {\bibinfo {author} {\bibfnamefont {E.}~\bibnamefont
  {Barausse}}, \bibinfo {author} {\bibfnamefont {R.}~\bibnamefont {Brito}},
  \bibinfo {author} {\bibfnamefont {V.}~\bibnamefont {Cardoso}}, \bibinfo
  {author} {\bibfnamefont {I.}~\bibnamefont {Dvorkin}}, \ and\ \bibinfo
  {author} {\bibfnamefont {P.}~\bibnamefont {Pani}},\ }\href {\doibase
  10.1088/1361-6382/aae1de} {\bibfield  {journal} {\bibinfo  {journal} {Class.
  Quant. Grav.}\ }\textbf {\bibinfo {volume} {35}},\ \bibinfo {pages} {20LT01}
  (\bibinfo {year} {2018})},\ \Eprint {http://arxiv.org/abs/1805.08229}
  {arXiv:1805.08229 [gr-qc]} \BibitemShut {NoStop}%
\bibitem [{\citenamefont {Mastrogiovanni}\ \emph {et~al.}(2025)\citenamefont
  {Mastrogiovanni}, \citenamefont {Maggio},\ and\ \citenamefont
  {Mascioli}}]{Mastrogiovanni:2025ixe}%
  \BibitemOpen
  \bibfield  {author} {\bibinfo {author} {\bibfnamefont {S.}~\bibnamefont
  {Mastrogiovanni}}, \bibinfo {author} {\bibfnamefont {E.}~\bibnamefont
  {Maggio}}, \ and\ \bibinfo {author} {\bibfnamefont {A.~F.}\ \bibnamefont
  {Mascioli}},\ }\href@noop {} {\  (\bibinfo {year} {2025})},\ \Eprint
  {http://arxiv.org/abs/2502.07675} {arXiv:2502.07675 [gr-qc]} \BibitemShut
  {NoStop}%
\bibitem [{\citenamefont {Fan}\ and\ \citenamefont {Chen}(2018)}]{Fan:2017cfw}%
  \BibitemOpen
  \bibfield  {author} {\bibinfo {author} {\bibfnamefont {X.-L.}\ \bibnamefont
  {Fan}}\ and\ \bibinfo {author} {\bibfnamefont {Y.-B.}\ \bibnamefont {Chen}},\
  }\href {\doibase 10.1103/PhysRevD.98.044020} {\bibfield  {journal} {\bibinfo
  {journal} {Phys. Rev. D}\ }\textbf {\bibinfo {volume} {98}},\ \bibinfo
  {pages} {044020} (\bibinfo {year} {2018})},\ \Eprint
  {http://arxiv.org/abs/1712.00784} {arXiv:1712.00784 [gr-qc]} \BibitemShut
  {NoStop}%
\bibitem [{Note1()}]{Note1}%
  \BibitemOpen
  \bibinfo {note} {{{The derivative self-interactions descend from the
  Lagrangian $\protect \mathcal {L}=(1-\alpha |\Psi |^2)|\partial \Psi |^2$ at
  leading order in $\alpha |\Psi |^2\ll 1$.}}}\BibitemShut {Stop}%
\bibitem [{\citenamefont {Comins}\ and\ \citenamefont
  {Schutz}()}]{Comins:1978}%
  \BibitemOpen
  \bibfield  {author} {\bibinfo {author} {\bibfnamefont {N.}~\bibnamefont
  {Comins}}\ and\ \bibinfo {author} {\bibfnamefont {B.~F.}\ \bibnamefont
  {Schutz}},\ }\href {\doibase 10.1098/rspa.1978.0196} {\bibfield  {journal}
  {\bibinfo  {journal} {Proc. R. Soc. Lond.}\ }\textbf {\bibinfo {volume}
  {364}},\ 10.1098/rspa.1978.0196}\BibitemShut {NoStop}%
\bibitem [{\citenamefont {{Hartle}}(1967)}]{Hartle1967}%
  \BibitemOpen
  \bibfield  {author} {\bibinfo {author} {\bibfnamefont {J.~B.}\ \bibnamefont
  {{Hartle}}},\ }\href {\doibase 10.1086/149400} {\bibfield  {journal}
  {\bibinfo  {journal} {Astrophys. J.}\ }\textbf {\bibinfo {volume} {150}},\
  \bibinfo {pages} {1005} (\bibinfo {year} {1967})}\BibitemShut {NoStop}%
\bibitem [{\citenamefont {Kleihaus}\ \emph {et~al.}(2005)\citenamefont
  {Kleihaus}, \citenamefont {Kunz},\ and\ \citenamefont
  {List}}]{Kleihaus:2005me}%
  \BibitemOpen
  \bibfield  {author} {\bibinfo {author} {\bibfnamefont {B.}~\bibnamefont
  {Kleihaus}}, \bibinfo {author} {\bibfnamefont {J.}~\bibnamefont {Kunz}}, \
  and\ \bibinfo {author} {\bibfnamefont {M.}~\bibnamefont {List}},\ }\href
  {\doibase 10.1103/PhysRevD.72.064002} {\bibfield  {journal} {\bibinfo
  {journal} {Phys. Rev. D}\ }\textbf {\bibinfo {volume} {72}},\ \bibinfo
  {pages} {064002} (\bibinfo {year} {2005})},\ \Eprint
  {http://arxiv.org/abs/gr-qc/0505143} {arXiv:gr-qc/0505143} \BibitemShut
  {NoStop}%
\bibitem [{\citenamefont {Siemonsen}\ and\ \citenamefont
  {East}(2021)}]{Siemonsen:2020hcg}%
  \BibitemOpen
  \bibfield  {author} {\bibinfo {author} {\bibfnamefont {N.}~\bibnamefont
  {Siemonsen}}\ and\ \bibinfo {author} {\bibfnamefont {W.~E.}\ \bibnamefont
  {East}},\ }\href {\doibase 10.1103/PhysRevD.103.044022} {\bibfield  {journal}
  {\bibinfo  {journal} {Phys. Rev. D}\ }\textbf {\bibinfo {volume} {103}},\
  \bibinfo {pages} {044022} (\bibinfo {year} {2021})},\ \Eprint
  {http://arxiv.org/abs/2011.08247} {arXiv:2011.08247 [gr-qc]} \BibitemShut
  {NoStop}%
\bibitem [{\citenamefont {Cardoso}\ \emph {et~al.}(2008)\citenamefont
  {Cardoso}, \citenamefont {Pani}, \citenamefont {Cadoni},\ and\ \citenamefont
  {Cavaglia}}]{Cardoso:2007az}%
  \BibitemOpen
  \bibfield  {author} {\bibinfo {author} {\bibfnamefont {V.}~\bibnamefont
  {Cardoso}}, \bibinfo {author} {\bibfnamefont {P.}~\bibnamefont {Pani}},
  \bibinfo {author} {\bibfnamefont {M.}~\bibnamefont {Cadoni}}, \ and\ \bibinfo
  {author} {\bibfnamefont {M.}~\bibnamefont {Cavaglia}},\ }\href {\doibase
  10.1103/PhysRevD.77.124044} {\bibfield  {journal} {\bibinfo  {journal} {Phys.
  Rev. D}\ }\textbf {\bibinfo {volume} {77}},\ \bibinfo {pages} {124044}
  (\bibinfo {year} {2008})},\ \Eprint {http://arxiv.org/abs/0709.0532}
  {arXiv:0709.0532 [gr-qc]} \BibitemShut {NoStop}%
\bibitem [{\citenamefont {{See Supplemental Material for more
  details}}()}]{sm}%
  \BibitemOpen
  \bibfield  {author} {\bibinfo {author} {\bibnamefont {{See Supplemental
  Material for more details}}},\ }\href@noop {} {}\BibitemShut {NoStop}%
\bibitem [{\citenamefont {John}(1981)}]{fritz_1981}%
  \BibitemOpen
  \bibfield  {author} {\bibinfo {author} {\bibfnamefont {F.}~\bibnamefont
  {John}},\ }\href {\doibase https://doi.org/10.1002/cpa.3160340103} {\bibfield
   {journal} {\bibinfo  {journal} {Communications on Pure and Applied
  Mathematics}\ }\textbf {\bibinfo {volume} {34}},\ \bibinfo {pages} {29}
  (\bibinfo {year} {1981})},\ \Eprint
  {http://arxiv.org/abs/https://onlinelibrary.wiley.com/doi/pdf/10.1002/cpa.3160340103}
  {https://onlinelibrary.wiley.com/doi/pdf/10.1002/cpa.3160340103} \BibitemShut
  {NoStop}%
\bibitem [{Note2()}]{Note2}%
  \BibitemOpen
  \bibinfo {note} {Specifically, $A_{\ell \ell }\sim \kappa ^{(p+q)/2} M^{p+q}
  A_{11}^p A_{22}^q$ and $A_{\ell \ell }\sim \alpha ^{(p'+q')/2} A_{11}^{p'}
  A_{22}^{q'}$, where $p,q,p',q'>0$ grow roughly linearly with $\ell
  $.}\BibitemShut {Stop}%
\bibitem [{\citenamefont {Baryakhtar}\ \emph {et~al.}(2021)\citenamefont
  {Baryakhtar}, \citenamefont {Galanis}, \citenamefont {Lasenby},\ and\
  \citenamefont {Simon}}]{Baryakhtar:2020gao}%
  \BibitemOpen
  \bibfield  {author} {\bibinfo {author} {\bibfnamefont {M.}~\bibnamefont
  {Baryakhtar}}, \bibinfo {author} {\bibfnamefont {M.}~\bibnamefont {Galanis}},
  \bibinfo {author} {\bibfnamefont {R.}~\bibnamefont {Lasenby}}, \ and\
  \bibinfo {author} {\bibfnamefont {O.}~\bibnamefont {Simon}},\ }\href
  {\doibase 10.1103/PhysRevD.103.095019} {\bibfield  {journal} {\bibinfo
  {journal} {Phys. Rev. D}\ }\textbf {\bibinfo {volume} {103}},\ \bibinfo
  {pages} {095019} (\bibinfo {year} {2021})},\ \Eprint
  {http://arxiv.org/abs/2011.11646} {arXiv:2011.11646 [hep-ph]} \BibitemShut
  {NoStop}%
\bibitem [{Note3()}]{Note3}%
  \BibitemOpen
  \bibinfo {note} {The $\ell =m=1$ mode undergoes the same instability, since
  the growth rates of $-\langle E_{11}\rangle $ for $t/\tau ^{\protect \rm
  EI}_{m=1}\lesssim 4$ and around $t/\tau ^{\protect \rm EI}_{m=1}\gtrsim 35$
  are identical; see Fig.~\ref {fig:mode_saturation}.}\BibitemShut {Stop}%
\bibitem [{Note4()}]{Note4}%
  \BibitemOpen
  \bibinfo {note} {We restrict $\tau _\ell ^{\protect \rm pNL}$ (as defined in
  the caption of Fig.~\ref {fig:nl_timescales}) to modes with $\ell \leq 5$,
  since after the second unstable phase (around $t/\tau ^{\protect \rm
  EI}_{m=1}\sim 55$ in Fig.~\ref {fig:mode_saturation}), it is less clear if
  only immediately neighboring modes are relevant for energy
  transfer.}\BibitemShut {Stop}%
\bibitem [{Note5()}]{Note5}%
  \BibitemOpen
  \bibinfo {note} {This suggests that the turbulent process is inefficiently
  triggered by off-resonance energy transfer. In a system with discrete linear
  frequencies, such as the one considered here, the resonance condition $\omega
  _{\protect \bar {m}}+\omega _{m''}=\omega _m+\omega _{m'}$ is violated
  leading to a slow-down of turbulent cascades (see e.g., Refs.~\cite
  {zakharov2012kolmogorov,nazarenko}).}\BibitemShut {Stop}%
\bibitem [{\citenamefont {Baryakhtar}\ \emph {et~al.}(2017)\citenamefont
  {Baryakhtar}, \citenamefont {Lasenby},\ and\ \citenamefont
  {Teo}}]{Baryakhtar:2017ngi}%
  \BibitemOpen
  \bibfield  {author} {\bibinfo {author} {\bibfnamefont {M.}~\bibnamefont
  {Baryakhtar}}, \bibinfo {author} {\bibfnamefont {R.}~\bibnamefont {Lasenby}},
  \ and\ \bibinfo {author} {\bibfnamefont {M.}~\bibnamefont {Teo}},\ }\href
  {\doibase 10.1103/PhysRevD.96.035019} {\bibfield  {journal} {\bibinfo
  {journal} {Phys. Rev. D}\ }\textbf {\bibinfo {volume} {96}},\ \bibinfo
  {pages} {035019} (\bibinfo {year} {2017})},\ \Eprint
  {http://arxiv.org/abs/1704.05081} {arXiv:1704.05081 [hep-ph]} \BibitemShut
  {NoStop}%
\bibitem [{\citenamefont {May}\ \emph {et~al.}(2025)\citenamefont {May},
  \citenamefont {East},\ and\ \citenamefont {Siemonsen}}]{May:2024npn}%
  \BibitemOpen
  \bibfield  {author} {\bibinfo {author} {\bibfnamefont {T.}~\bibnamefont
  {May}}, \bibinfo {author} {\bibfnamefont {W.~E.}\ \bibnamefont {East}}, \
  and\ \bibinfo {author} {\bibfnamefont {N.}~\bibnamefont {Siemonsen}},\ }\href
  {\doibase 10.1103/PhysRevD.111.044062} {\bibfield  {journal} {\bibinfo
  {journal} {Phys. Rev. D}\ }\textbf {\bibinfo {volume} {111}},\ \bibinfo
  {pages} {044062} (\bibinfo {year} {2025})},\ \Eprint
  {http://arxiv.org/abs/2410.21442} {arXiv:2410.21442 [gr-qc]} \BibitemShut
  {NoStop}%
\bibitem [{Note6()}]{Note6}%
  \BibitemOpen
  \bibinfo {note} {In the eikonal limit $\omega _{m=\ell }\rightarrow m\omega
  _-$, and the resonance condition $\omega _{\protect \bar {m}}+\omega
  _m=\omega _{m'}+\omega _{m''}$ is satisfying asymptotically.}\BibitemShut
  {Stop}%
\bibitem [{\citenamefont {Boerner}\ \emph {et~al.}(2023)\citenamefont
  {Boerner}, \citenamefont {Deems}, \citenamefont {Furlani}, \citenamefont
  {Knuth},\ and\ \citenamefont {Towns}}]{access}%
  \BibitemOpen
  \bibfield  {author} {\bibinfo {author} {\bibfnamefont {T.~J.}\ \bibnamefont
  {Boerner}}, \bibinfo {author} {\bibfnamefont {S.}~\bibnamefont {Deems}},
  \bibinfo {author} {\bibfnamefont {T.~R.}\ \bibnamefont {Furlani}}, \bibinfo
  {author} {\bibfnamefont {S.~L.}\ \bibnamefont {Knuth}}, \ and\ \bibinfo
  {author} {\bibfnamefont {J.}~\bibnamefont {Towns}},\ }in\ \href {\doibase
  10.1145/3569951.3597559} {\emph {\bibinfo {booktitle} {Practice and
  Experience in Advanced Research Computing 2023: Computing for the Common
  Good}}},\ \bibinfo {series and number} {PEARC '23}\ (\bibinfo  {publisher}
  {Association for Computing Machinery},\ \bibinfo {address} {New York, NY,
  USA},\ \bibinfo {year} {2023})\ pp.\ \bibinfo {pages} {173--176}\BibitemShut
  {NoStop}%
\bibitem [{Note7()}]{Note7}%
  \BibitemOpen
  \bibinfo {note} {Single-mode initial data sources all odd-$\protect \bar
  {\ell }=\protect \bar {m}$ modes if $\alpha \protect \neq 0$.}\BibitemShut
  {Stop}%
\bibitem [{Note8()}]{Note8}%
  \BibitemOpen
  \bibinfo {note} {There may still be an inverse cascade, when considering
  initial data with support only across high-$\ell $ modes at fixed
  $m$.}\BibitemShut {Stop}%
\bibitem [{Note9()}]{Note9}%
  \BibitemOpen
  \bibinfo {note} {Note, the definition of $\Omega $ differs from the one used
  in Ref.~\cite {Siemonsen:2020hcg}.}\BibitemShut {Stop}%
\end{thebibliography}%

\appendix
\section{Details on the scalar theory} \label{app:theory}

In the following, we provide further details on the nonlinear scalar theory outlined in the main text. To define the coupling coefficients $N^{\bar{m}mm'm''}_{\bar{\ell}\ell\ell'\ell''}(\Psi)$ used in the main text, we being by introducing 
\begin{align}
D^{\bar{m}mm'm''}_{\bar{\ell}\ell\ell'\ell''} \ & =\int_{S^2} d\Omega \ Y^*_{\bar{\ell}\bar{m}}Y_{\ell m}Y^*_{\ell'' m''}Y_{\ell'm'}.
\end{align}
This appears, when projecting the scalar self-interaction $\Psi |\Psi|^2$ onto the spherical harmonic mode $(\bar{\ell},\bar{m})$ and can be integrated explicitly to
\begin{align}
\begin{aligned}
 & \ D^{\bar{m}mm'm''}_{\bar{\ell}\ell\ell'\ell''} = (-1)^{m''+m}C_{\bar{\ell}\ell\ell'\ell''} \\
& \ \qquad \times \sum_j\frac{2j+1}{4\pi}
\begin{pmatrix}
\bar{\ell} & \ell & j \\
0 & 0 & 0
\end{pmatrix}
\begin{pmatrix}
\ell' & \ell'' & j \\
0 & 0 & 0
\end{pmatrix}
\\
& \ \qquad \times
\begin{pmatrix}
\bar{\ell} & \ell & j \\
-\bar{m} & m & m'-m''
\end{pmatrix}
\begin{pmatrix}
\ell' & \ell'' & j \\
m' & -m'' & m-\bar{m}
\end{pmatrix},
\end{aligned}
\label{eq:Dllll}
\end{align}
using $C_{\bar{\ell}\ell\ell'\ell''}=\sqrt{2\bar{\ell}+1}\sqrt{2\ell+1}\sqrt{2\ell'+1}\sqrt{2\ell''+1}$ and $(\cdot)$ denoting the Wigner 3-j symbol. For the derivative self-interactions, the coefficients
\begin{align}
E^{\bar{m}mm'm''}_{\bar{\ell}\ell\ell'\ell''}=\int d\Omega \frac{Y^*_{\bar{\ell}\bar{m}}Y_{\ell m} Y_{\ell' m'}Y^*_{\ell'' m''}}{\sin^2\theta},
\label{eq:integralE}
\end{align}
are needed. These coefficients and $D^{\bar{m}mm'm''}_{\bar{\ell}\ell\ell'\ell''}$ enter the nonlinear couplings
\begin{widetext}
\begin{align}
\begin{aligned}
N^{\bar{m}m m' m''}_{\bar{\ell}\ell\ell'\ell''}(\Psi)=\ & \frac{\kappa f}{r^2}\phi_{\ell m}\phi_{\ell'm'}\phi^*_{\ell''m''} D^{\bar{m}m m' m''}_{\bar{\ell}\ell\ell'\ell''}+\frac{\alpha f^2}{l r^4}\Bigg[D^{\bar{m}m m' m''}_{\bar{\ell}\ell\ell'\ell''}\Big(\frac{l}{f^2}(m\Omega\phi_{\ell m}-ir\dot{\phi}_{\ell m})(m'\Omega\phi_{\ell' m'}-ir\dot{\phi}_{\ell' m'}) \\
& +(\phi_{\ell m}-r\partial_r\phi_{\ell m})(\phi_{\ell' m'}-r\partial_r\phi_{\ell' m'}) \Big)\phi^*_{\ell'' m''}+\phi_{\ell m}\phi_{\ell' m'}\phi^*_{\ell'' m''}\Big( -mm'E^{\bar{m}m m' m''}_{\bar{\ell}\ell\ell'\ell''} \\
& + \tilde{c}_{++}E^{\bar{m}m m' m''}_{\bar{\ell},\ell+1,\ell'+1,\ell''}+\tilde{c}_{+-}E^{\bar{m}m m' m''}_{\bar{\ell},\ell+1,\ell'-1,\ell''}+\tilde{c}_{-+}E^{\bar{m}m m' m''}_{\bar{\ell},\ell-1,\ell'+1,\ell''}+\tilde{c}_{--}E^{\bar{m}m m' m''}_{\bar{\ell},\ell-1,\ell'-1,\ell''} \Big)\Bigg].
\end{aligned}
\label{eq:N}
\end{align}
Here we used 
\begin{align}
\tilde{c}_{++}= \ & \ell\ell'\left[\frac{(1+\ell-m)(1+\ell+m)(1+\ell'+m')(1+\ell'-m')}{(3+4\ell(2+\ell))(3+4\ell'(2+\ell'))}\right]^{1/2},\\
\tilde{c}_{-+}= \ & -\ell'(\ell+1)\left[\frac{(\ell+m)(1+\ell'+m')(1+\ell'-m')(\ell-m)}{(3+2\ell')(4\ell^2-1)(2\ell'+1)}\right]^{1/2},\\
\tilde{c}_{--}= \ & (\ell+1)(\ell'+1)\left[\frac{(\ell^2-m^2)(\ell'^2-m'^2)}{(4\ell'^2-1)(4\ell^2-1)}\right]^{1/2},
\end{align}
where $\tilde{c}_{+-}= \tilde{c}_{-+} \ (\ell \leftrightarrow \ell',m \leftrightarrow m')$. Note, for $\ell=\ell'=0$ all $\tilde{c}$ coefficients vanish. 
\end{widetext}

The integral \eqref{eq:integralE} is divergent for $\bar{m}=m=m'=m''=0$, independently of the polar mode numbers. This is irrelevant for the term $\sim mm'E^{\bar{m}m m' m''}_{\bar{\ell}\ell\ell'\ell''}$, and when assuming $\bar{\ell},\ell,\ell',\ell''=0$, also for the last line in eq.~\eqref{eq:N}. While not manifestly finite, the divergences cancel exactly in the last line of eq.~\eqref{eq:N} also for $\bar{\ell},\ell,\ell',\ell''>0$. Regardless, all solutions of interest in this work are obtained neglecting the higher-order polar modes for $\bar{m}=m=m'=m''=0$, therefore, circumventing any formally divergent terms in the above expression for $N^{\bar{m}m m' m''}_{\bar{\ell}\ell\ell'\ell''}(\Psi)$.

\begin{figure*}[t!]
\includegraphics[width=0.32\linewidth]{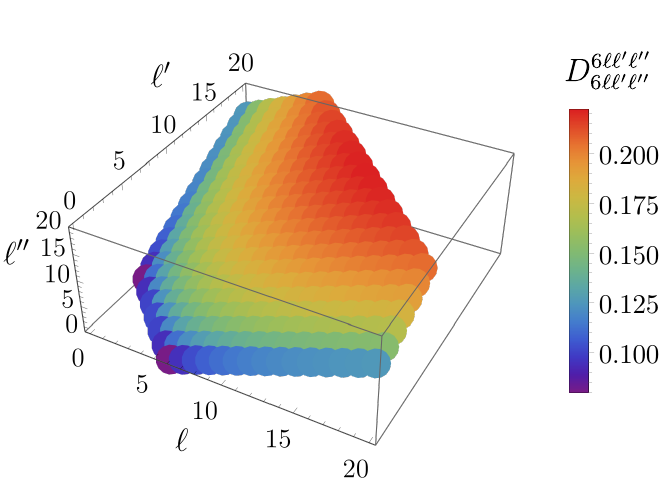}\hfill
\includegraphics[width=0.32\linewidth]{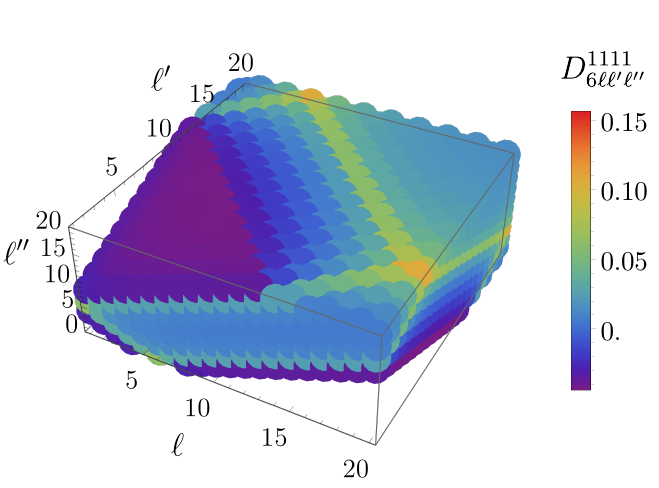}\hfill
\includegraphics[width=0.32\linewidth]{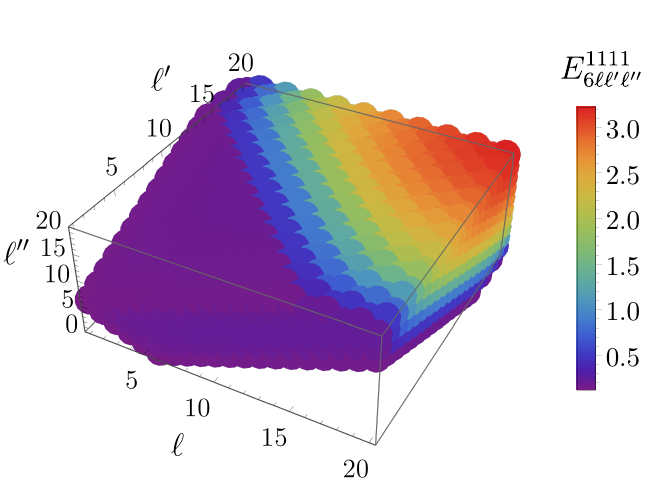}
\caption{(left) The quantity $D_{6\ell\ell'\ell''}^{6\ell\ell'\ell''}$ plotted over the index space $(\ell,\ell',\ell'')$ up to a maximum mode number of $\ell,\ell',\ell''\leq 20$. Color indicates its value. For clarity, when the coupling coefficients vanish, we remove the dots. While the value of $E_{6\ell\ell'\ell''}^{6\ell\ell'\ell''}$ differs slightly, they are qualitatively similar. (center and right) The coupling coefficients at fixed $\bar{m}=m=m'=m''=1$, i.e., $D^{1111}_{6\ell\ell'\ell''}$ (center) and $E^{1111}_{6\ell\ell'\ell''}$ (right). In the eikonal limit, these coefficients behave as $E^{1111}_{1\ell\ell\ell}\sim\ell^{1/2}$ (for odd-$\ell$ modes), whereas $D^{1111}_{1\ell\ell\ell}\sim\ell^{-3/2}$.}
\label{fig:indexspace}
\end{figure*}

Let us gain intuition for the coefficients $D^{\bar{m}mm'm''}_{\bar{\ell}\ell\ell'\ell''}$ and $E^{\bar{m}mm'm''}_{\bar{\ell}\ell\ell'\ell''}$. When focussing on the coupling between the fastest growing ergoregion unstable modes, i.e., $\ell=m$ etc., then these coefficients reduce to $D^{\bar{\ell}\ell\ell'\ell''}_{\bar{\ell}\ell\ell'\ell''}$ and $E^{\bar{\ell}\ell\ell'\ell''}_{\bar{\ell}\ell\ell'\ell''}$, which are non-vanishing if $\bar{\ell}+\ell''=\ell+\ell'$. This resonance condition corresponds to a plane in three-dimensional index-space for a given mode $\bar{\ell}$. As an example, in Fig.~\ref{fig:indexspace}, we show those non-vanishing coefficients sourcing the $\bar{\ell}=6$ mode. As can be seen from there, the coefficients only weakly depend on the index set $(\ell,\ell',\ell'')$, with $D^{6\ell\ell'\ell''}_{6\ell\ell'\ell''}\sim \mathcal{O}(0.1)$. It also follows form this, if $\alpha = 0$ two-mode initial data (as chosen in this work) is required to activate the non-trivial couplings induced by $D^{\bar{\ell}\ell\ell'\ell''}_{\bar{\ell}\ell\ell'\ell''}$ between the $\ell=m$ modes\footnote{Single-mode initial data sources all odd-$\bar{\ell}=\bar{m}$ modes if $\alpha\neq 0$.}. If $\alpha=0$, but $\kappa>0$, initial data with support only over the $\ell=m=1$ mode leads to no mixing into other $\ell=m$ modes, and induces only a single mode-level self-interaction through $D^{1111}_{1111}=3/(10\pi)$; this results in a smooth (non-turbulent) saturation of the $\ell=m=1$ instability from such single-mode initial data.

Away from the set of $\ell=m$ modes, below we are interested also in the behavior of the system in the nonlinear regime, when keeping $m$ fixed, but considering modes with $\ell\geq m$. In Fig.~\ref{fig:indexspace}, we show the non-vanishing couplings between the lowest $\ell$-modes at fixed azimuthal index $m=1$. In this case, the non-vanishing coefficients no longer form a simple plane in index space, but rather a band with width $>1$. The coefficients $D^{mmmm}_{\bar{\ell}\ell\ell'\ell''}$ share these features for other fixed azimuthal indices $m\neq 1$ as well. In the eikonal limit $E^{1111}_{\bar{\ell}\ell\ell\ell}\sim \ell^{1/2}$ for fixed $\bar{\ell}$, while $D^{1111}_{\bar{\ell}\ell\ell\ell}\sim \ell^{-3/2}$. This is a signature of the scale-dependence of the derivative self-interactions compared with their potential-type counterparts.

As noted in the main text, our truncated theory with derivative self-interactions exhibits only a perturbatively conserved energy for $\alpha|\Psi|^2\ll 1$. Therefore, we focus here entirely on the conservation of the case with $\alpha=0$. In this setting, the stress-energy tensor associated with the nonlinear scalar theory considered in the main text is explicitly given by
\begin{align}
T_{\mu\nu}=2\partial_{(\mu}\Psi^*\partial_{\nu)}\Psi-g_{\mu\nu}\left(g^{\alpha\beta}\partial_{(\alpha}\Psi^*\partial_{\beta)}\Psi+\frac{\kappa}{2}|\Psi|^{4}\right).
\end{align}
Due to the presence of an asymptotically timelike Killing vector field $\xi^\mu$, the current $j^\mu=-T^{\mu\nu}\xi_\nu$ is convered: $\nabla_\mu j^\mu=0$. This implies that the total energy $E[\Psi]=-\int_V d^3x\sqrt{-g} T^t_t$ is conserved in volume $V\subset\Sigma_t$ within each $t=\text{const.}$ slice $\Sigma_t$ of the spacetime in the sense that $\partial_tE[\Psi]=-\mathcal{F}$; here $\mathcal{F}$ is the outgoing scalar energy flux at the boundary $\partial V$: $\mathcal{F}=\int_{\partial V} d^2\sigma_\mu T^{\mu\nu}\xi_\nu$. With the ansatz $\Psi=r^{-1}\sum_I \phi_{\ell m}(t,r)Y_{\ell m}(\theta,\varphi)$, the total energy splits as $E[\Psi]=E_{\rm int}+\sum_I E_{\ell m}$ into the linear components
\begin{align}
\begin{aligned}
E_{\ell m} = \int dr \frac{l^{3/2}}{f} \bigg[\frac{f}{l}\frac{|\phi_{\ell m}|^2}{r^2}\Big(\ell(\ell+1)+1\Big) -\frac{|\phi_{\ell m}|^2m^2\Omega^2}{f} \\
+\frac{|\partial_t\phi_{\ell m}|^2}{f} -\frac{f(\phi_{\ell m}^*\partial_r\phi_{\ell m}+\phi_{\ell m}\partial_r \phi^*_{\ell m})}{lr} +\frac{f|\partial_r \phi_{\ell m}|^2}{l}\bigg],
\end{aligned}
\end{align}
as well as the interaction energy
\begin{align}
E_{\rm int}=\int dr\frac{\kappa l^{3/2}}{2r^2f}\sum_{\bar{I},I,I',I''}D^{\bar{m}mm'm''}_{\bar{\ell}\ell \ell'\ell''} \phi^*_{\bar{\ell}\bar{m}}\phi_{\ell m}\phi_{\ell' m'}\phi^*_{\ell'' m''}.
\label{eq:Eint}
\end{align}
Similarly, the total outgoing scalar energy flux splits into $\mathcal{F}=\sum_I \mathcal{F}_{\ell m}$.
Note, for $\alpha,\kappa=0$ the linear components are all conserved individually, with associated fluxes: $\partial_t E_{\ell m}=-\mathcal{F}_{\ell m}$. We also compute outgoing angular momentum fluxes $\mathcal{J}_{\ell m}$ in direct analogy to this.

\red{\section{Details on nonlinear energy evolution}

\begin{figure}
\includegraphics[width=1\linewidth]{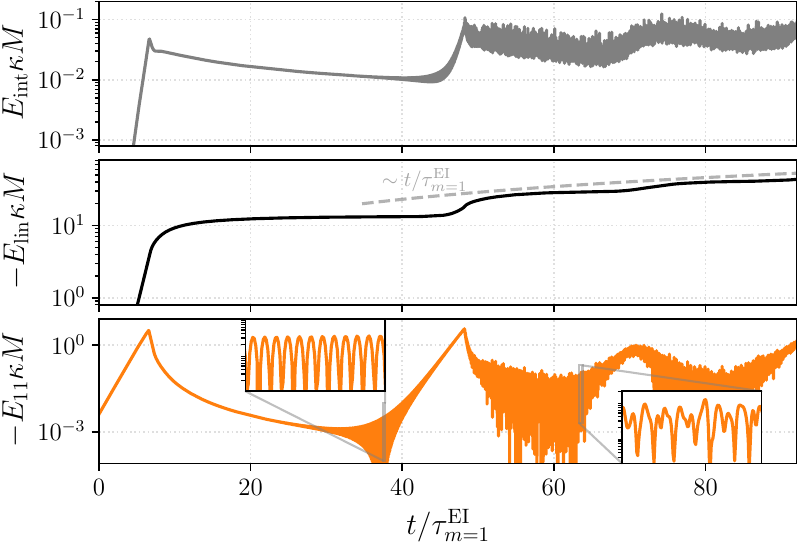}
\caption{The evolution of the interaction energy $E_{\rm int}$, the sum of linear energies $E_{\rm lin}$, and the linear energy $E_{11}$, throughout saturation of the system, when considering only potential-type self-interactions.}
\label{fig:total_energy}
\end{figure}

Finally, for potential-type self-interactions we show the evolution of the interaction energy, the sum of linear energies, and the linear energy of the $\ell=m=1$ mode in Fig.~\ref{fig:total_energy}. First, this confirms the weakly nonlinear nature of the turbulent cascade, since $E_{\rm int}\ll E_{\rm lin}$. Secondly, self-interactions bound the energy in the most unstable mode, $|E_{11}|$, from above, while the ergoregion instability effectively bounds it from below. This, together with the energy transport mechanism to smaller scales, implies a on long timescales constant (negative) energy injection rate into the $\ell=m=1$ mode, and hence, a linear-in-time growth of the total energy $|E[\Psi]|\approx|E_{\rm lin}|$, as indicated in the middle panel of Fig.~\ref{fig:total_energy}. Lastly, during the subsequent exponential growth and saturation of the $\ell=m=1$ mode, i.e., for $t/\tau^{\rm EI}_{m=1}\gtrsim 55$, all mode energies exhibit modulations (shown in the insets of the bottom panel of Fig.~\ref{fig:total_energy}) on timescales comparable to or larger than $1/\text{Re}(\omega_{m=\ell})$, but shorter than $\tau_\ell^{\rm pNL}$. This may be explained as follows: in contrast to the initial saturation, where the amplitude of high-$\ell$ modes is exponentially suppressed, during the second (and any subsequent) linearly unstable growth of the $\ell=m=1$ mode, there exists a sector of strongly coupled modes centered around $\ell_c=7$, with amplitudes comparable to that of the $\ell=m=1$ mode (see the mode spectra shown in the main Letter). These constitute sources for the low-$\ell$ modes with driving frequency comparable to, but different from, their linear frequencies, leading to modulations of their energies. For instance, the combination $\phi_{55}^2\phi_{99}^*$ drives the $\ell=m=1$ mode at frequency $\omega_{\rm source}=\text{Re}(2\omega_5-\omega_9)$, which leads to modulations with timescale $1/[\omega_{\rm source}-\text{Re}(\omega_{11})]$ matching the oscillations in Fig.~\ref{fig:total_energy}. This modulation was averaged over for clarity in the energy evolutions shown in the main Letter.}

\section{Influence of other modes} \label{app:mode_dep}

\subsubsection{Potential-type self-interactions}

\begin{figure}
\includegraphics[width=1\linewidth]{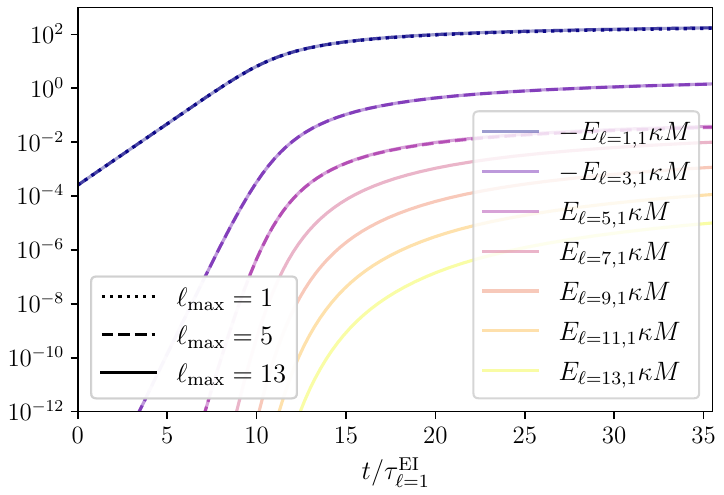}
\caption{The evolution of the linear energies $E_{\ell m}$ through the ergoregion instability (of the $\ell=m=1$ mode) and subsequent saturation, when considering only couplings with azimuthal index $m=1$, but (potentially large) polar modes $\ell\geq 1$. Here, $\ell_{\max}=1,5,13$ indicates three different numerical evolutions, which include polar modes only up to (and including) $\ell_{\max}$. Recall, only a finite number of energies $E_{\ell m}$ can be negative, as only a finite number of $\ell$-modes are ergoregion unstable at fixed $m$ \cite{Siemonsen:2025wib}. The dynamics of $-E_{\ell=1,1}\kappa M$ are practically independent of $\ell_{\max}$.}
\label{fig:polar_sat}
\end{figure}

In the following, the focus lies on potential-type self-interactions and mode couplings induced by $\kappa>0$, while assuming $\alpha=0$. We discuss both the importance of the monopolar mode in amplifying all higher-order azimuthal modes, as well as the role that other mode couplings play, beyond the $\ell=m\geq 0$ sector mainly discussed in the main text. To that end, we begin by considering the special case of fixed $\bar{m}=m=m'=m''=1$, but $\bar{\ell},\ell,\ell',\ell''\geq 1$ together with single-mode initial data, where $\phi_{11}$ and $\partial_t\phi_{11}$ are set to the corresponding fastest growing ergoregion unstable field configuration (i.e., zeroth radial overtone) with sufficiently small amplitude as to be deep in the linear regime; all other modes are initialized to zero. Under these assumptions, the scalar self-interactions \eqref{eq:Dllll} lead to mixing only in odd-$\ell$ modes. We vary $\ell_{\max}$ from $1$ to $13$ to investigate the dependence of the resulting dynamics on the dissipation scale.

In Fig.~\ref{fig:polar_sat}, the evolution of the linear energies $E_{\ell  1}$ through the saturation of the linear instability is shown. Focusing first on the $\ell_{\max}=13$ case, we find that the nonlinear saturation of the instability exhibits no direct cascade towards higher-$\ell$ modes\footnote{There may still be an inverse cascade, when considering initial data with support only across high-$\ell$ modes at fixed $m$.}; the saturation of the instability proceeds in a laminar fashion. In fact, even evolutions with $\ell_{\max}=1,5$ are practically identical to the case with $\ell_{\max}=13$. This shows that these polar couplings can be neglected during the saturation of the ergoregion instability of the $\ell=m=1$ mode. We have explicitly checked that initial data supported over all modes $\ell\geq m=1$ (i.e., including the even-$\ell$ modes) leads to qualitatively identical behavior, as the even-$\ell$ modes are not amplified during the nonlinear saturation of the instability.

\begin{figure}
\includegraphics[width=1\linewidth]{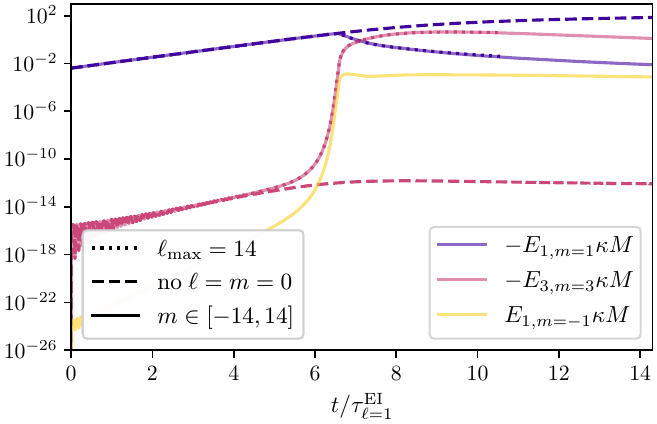}
\caption{Evolution of three select linear energies (different colors) through the nonlinear saturation of the ergoregion instability. Different line styles correspond to evolutions of the same initial data, but considering couplings between all $\ell=|m|$ modes with $m\in[-14,14]$ (solid), all modes with $\ell=m\in[0,14]$ (dotted), and all modes $\ell=m\in[1,14]$ (dashed).}
\label{fig:other_cases}
\end{figure}

Next we check, whether the inclusion of the $\ell=|m|$ modes with $m<0$ impacts the evolution. To that end, we construct the two-mode initial data considered from the main text (i.e., initialize $\phi_{11},\dot{\phi}_{11},\phi_{22},\dot{\phi}_{22}$ to their respective fastest growing ergoregion unstable states) and include all couplings between the modes $\ell=|m|$ with $m\in[-14,14]$ (any energy entering the $|m|=15$ modes is dissipated). In Fig.~\ref{fig:other_cases}, we compare the evolution of this system with that of the system, where only couplings between the $\ell=m\in[0,14]$ modes are included. From there, we conclude that the negative-$m$ and $\ell=|m|$ modes are sub-dominant and do not impact the qualitative evolution of the turbulent process during the nonlinear saturation. This justifies to only consider the $\ell=m\geq 0$ modes.

Finally, we also remove the $\ell=m=0$ mode (and associated couplings) from the evolution of the system and include only those with $\ell=m\geq 1$. This case is labeled ``no $\ell=m=0$'' in Fig.~\ref{fig:other_cases}. From there is becomes clear that this monopolar mode is pivotal for driving the amplification during the nonlinear saturation of the fastest growing ergoregion unstable state. That is, excluding the monopolar modes implies the system saturates in a laminar fashion qualitatively analogous to the case shown in Fig.~\ref{fig:polar_sat}. This suggests that the monopolar mode facilitates a parametric driving of higher azimuthal states similar to what was observed in Ref.~\cite{Baryakhtar:2020gao}. In particular, the decay rate of the $\ell=m=0$ mode is roughy one order of magnitude larger than the growth rate any of the ergoregion unstable modes; positive energy is transfered to the monopolar mode and efficiently dissipated, while amplifying the negative energy in the $\ell=m\geq 1$ states. In particular, as the monopolar mode drives the amplification of the $\ell=m=2$ mode, the $\ell=m=1$ state decays via the $(1,1)+(1,1)+(0,0)\rightarrow (2,2)$ process [where $(\ell,m)$ is the tuple of mode labels].

\begin{figure}
\includegraphics[width=1\linewidth]{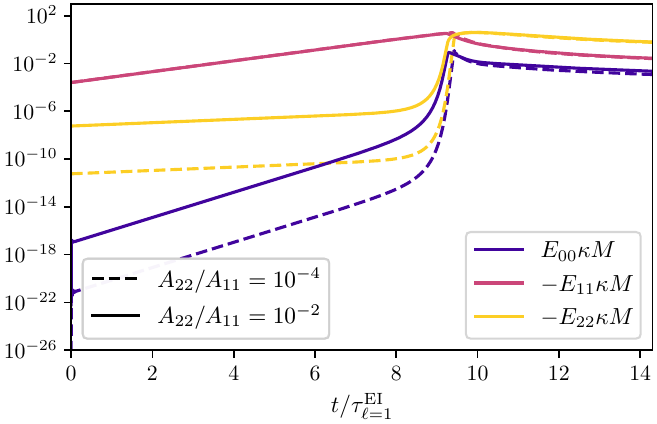}
\caption{Evolution of the system starting from the initial data described in the main text, but with two different initial amplitude ratios $A_{22}/A_{11}$. Note, the amplitude $A_{11}$ is a factor of four smaller than the one considered throughout this work (and shown e.g., in Fig.~\ref{fig:other_cases}). Only the first three linear energies $E_{\ell\ell}$ are presented.}
\label{fig:amp_ind}
\end{figure}

Lastly, we comment on the initial data dependence of the saturation dynamics. To that end, the focus lies only on the $\ell=m\geq 0$ modes and respective couplings. If only the $\phi_{11}$ and $\phi_{22}$ are non-vanishing initially, then we find only marginal dependence of the saturation dynamics, and in particular, the weakly turbulent cascade, on the ratio of amplitudes $A_{11}/A_{22}$. This is explicitly shown in Fig.~\ref{fig:amp_ind}. Since the monopolar drives a rapid amplification of all higher-order azimuthal modes, a weak dependence on the amplitude ratio is expected. We also checked that the evolution of the initial data, where all $\phi_{\ell\ell}$ were initially non-vanishing with the same amplitude, is qualitatively similar to that shown throughout this work. The nonlinear transfer timescales $\tau_\ell^{\rm NL}$ may, however, depend on the amplitude of these higher-order azimuthal modes; naturally, if a given such mode were initalized with amplitude away from zero, then the parameteric driving requires less to amplify this mode into the nonlinear regime. This has then minor impact on the time between peaks of the nonlinear energies, which impacts $\tau_\ell^{\rm NL}$.

\subsubsection{Derivative self-interactions} \label{app:int}

\begin{figure}
\includegraphics[width=1\linewidth]{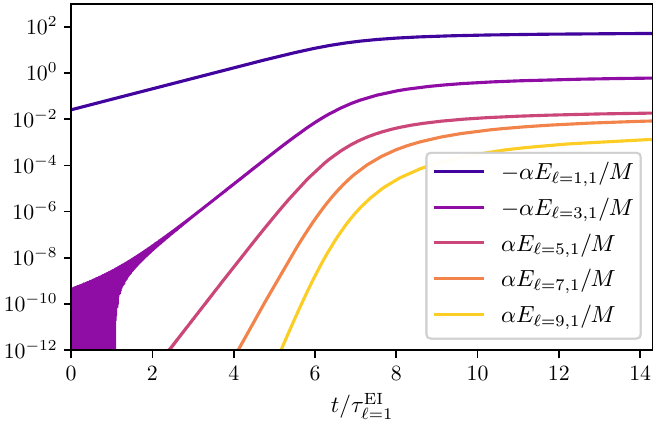}
\caption{Evolution of the higher-order polar modes (assuming $m=1$) given single-mode initial data and derivative scalar self-interactions. In this case $\ell_{\max}=10$.}
\label{fig:deriv}
\end{figure}

In addition to the discussion in the previous section in the context of potential-type self-interactions, we briefly turn to pure scalar derivative self-interactions: $\alpha>0$ and $\kappa=0$. To that end, we focus exclusively on the case of fixed azimuthal index $m=1$, but varying $\ell\geq 1$ (see Fig.~\ref{fig:indexspace} for the visualization of relevant coupling coefficients). As before, the initial data is set by the fastest growing erogregion unstable configuretion $\phi_{11}$ with amplitude deep in the linear regime. The evolution of these data is shown in Fig.~\ref{fig:deriv}. The evolution is qualitatively identical to the corresponding dynamics in the scalar theory with potential-type self-interactions (shown in Fig.~\ref{fig:polar_sat}). This is despite the scale-dependence of the derivative interactions, and their relevance for large $\ell$. In particular, there is no sign of energy transport to small scales, beyond exponentially suppressed mode mixing. 

\section{Details on the spinning boson star} \label{app:bs}

The spacetime considered in this work is that of a spinning bosons star in the solitonic scalar model with coupling constant $\sigma=0.2$ and azimuthal index $\tilde{m}=3$, as defined in Ref.~\cite{Siemonsen:2025wib}. This star solution is constructed using Newton-Raphson methods developed in Refs.~\cite{Kleihaus:2005me,Siemonsen:2020hcg}. In Lewis-Papapetrou coordinates, the associated metric takes the form\footnote{Note, the definition of $\Omega$ differs from the one used in Ref.~\cite{Siemonsen:2020hcg}.}
\begin{align}
\begin{aligned}
ds^2=-fdt^2+\frac{l}{f}\large\{g(dr^2+ \ & r^2d\theta^2) \\
+r^2 & \sin^2\theta (d\varphi-\Omega dt)^2 \large\},
\end{aligned}
\end{align}
where $f,l,g,\Omega$ all depend both on the radial and polar coordiantes. The slow-rotation approximation used throughout this work, and shown in the main text, is obtained from this by setting $g=1$ and evaluating $l,f,\Omega$ in the equatorial plane \cite{Siemonsen:2025wib,Cardoso:2007az}. The boson star serving as the background spacetime throughout this work has internal frequency $\tilde{\omega}/\mu=0.427$ (in units of scalar mass $\mu$), total mass $\mu M=3.85$, total angular momentum $J/M^2=1.013$, and compactness $M/R=0.46$ (where $R$ is the equatorial radius of the star~\cite{Siemonsen:2020hcg}).

\section{Numerical implementation} \label{app:numerics}

Here we briefly outline our numerical implementation to solve the set of equations presented in the main text. To effectively place resolution, we introduce the new radial coordinate
\begin{align}
r_{*}(r)=\int_0^r dr' \sqrt{\frac{l}{f^2}},
\label{eq:rstar}
\end{align}
and discretize the radial derivatives using fourth-order accurate finite differences across a uniform grid with spacing $\Delta r_*$ in this new coordinate. The time stepping is achieved by means of a fourth-order accurate Runge-Kutta method. The metric functions are obtained numerically with radial resolution $\Delta r_{\rm BS}$ coarser than our typical choices for $\Delta r_*$. The computational domain extends from $r_*=0$ up to $r=r_{\max}\geq 500 M$ (we have varied $r_{\max}$ to ensure independence of our results of this choice). At $r_{\max}$ and $r_*=0$ we impose outgoing Sommerfeld boundary and regularity conditions, respectively. Due to our choice of ansatz, $\Psi\sim \phi_{\ell m}(t,r)/r$, we found it useful to impose the regular behavior, $\phi_{\ell m}\sim r^\ell$ for $r\approx 0$, explicitly: the value of  $\phi(t=\text{const.},r_{*,i})$ at those grid points $r_{*,i}$ close to and at the origin, i.e., for $0\leq i\leq i_{\max}$, are obtained from an extrapolation of $\phi(t=\text{const.},r_{*,i})$ with $i>i_{\max}$ assuming $\phi_{\ell m}(t=\text{const.},r)=r^\ell (a+br^2+cr^4 +\mathcal{O}(r^6))$. By varying $i_{\max}$, we ensure that this choice has no impact on the results. Lastly, in all our evolutions we include only modes with polar mode number below the cutoff $\ell=\ell_{\max}+1$. The highest-$\ell$ mode, i.e., the $(\ell_{\max}+1)$-th mode, is artificially dissipated by setting $\phi_{\ell_{\max}+1,m}=0$ after each timestep for all relevant $m$. 

\begin{figure}
\includegraphics[width=1\linewidth]{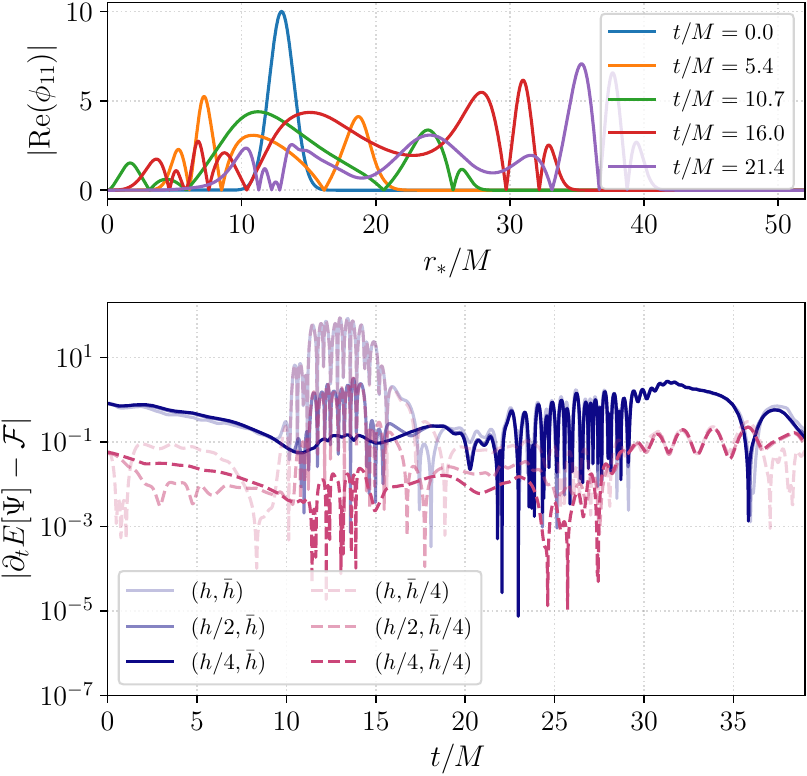}
\caption{The dynamics of $\phi_{11}$ from ingoing Gaussian-pulse initial data (top panel). The evolution of the corresponding energy conservation violations (bottom panel) for different reference grid spacing $h$ of the scalar evolution scheme, as well as $\bar{h}$ of the background's metric functions.}
\label{fig:large_conv}
\end{figure}

The implementation is tested in the linear and nonlinear regimes on the boson star background. We begin by considering large initial data, i.e., those in the strongly nonlinear regime. To that end, we focus on the potential-type self-interactions of the $\ell=m=1$ mode alone (i.e., ignoring all couplings to other modes) and initialize $\phi_{11}(0,r_*)=A \exp(-(r_*-\tilde{r})^2/\lambda^2), \dot{\phi}_{11}(0,r_*)=-2A\exp(-(r_*-\tilde{r})^2/\lambda^2)(r_*-\tilde{r})/\lambda^2$. We choose $\kappa=1/M^2$, $\tilde{r}=13M, \lambda=M$, and $A=10$, such that the system starts out in the strongly nonlinear regime, as $E_{\rm int}\sim |E_{\rm lin}|$. To investigate the convergence behavior of our implementation, we vary both the grid spacing $\Delta r_*$ and the resolution of the boson star metric functions $\Delta r_{\rm BS}$. In Fig.~\ref{fig:large_conv}, the violations of the energy conservation law, $|\partial_t E[\Psi]-\mathcal{F}|$, is used to measure the convergence behavior towards the continuum solution. The convergence of the solution inside and near the star depends on both the truncation error of the metric functions and that of the scalar evolution of the nonlinear scalar theory. Depending on where (in space) the perturbation resides at a given time, the numerical error is either dominated by the background's or the scalar evolution scheme truncation errors. This is likely, since the grids of the scalar evolution scheme's and the metric functions are non-uniform with respect to each other. Ultimately, the resolution of the background star solution is limited by the methods introduced in Ref.~\cite{Siemonsen:2020hcg}, so that for sufficiently high resolution of the scalar evolution scheme, the error (as defined above) ceases to converge to zero further.

\begin{figure}[t!]
\includegraphics[width=1\linewidth]{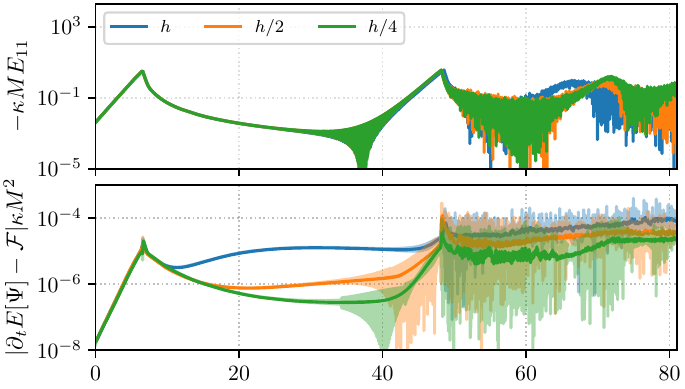}
\includegraphics[width=1\linewidth]{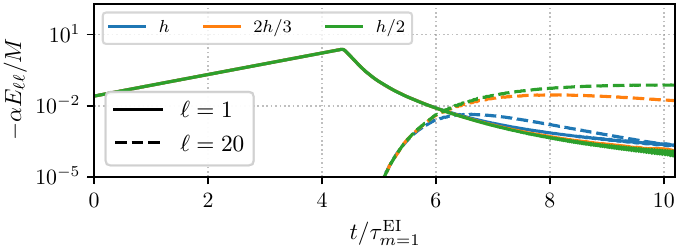}
\caption{Evolution of $E_{11}$ (top) and violations of the energy conservation (center) of the system going through saturation (assuming $\alpha=0$). We vary the radial resolution, with respect to the reference resolution $h$, of the scalar evolution scheme. Solid lines are moving time averages of the shaded lines in the central panel. (bottom) Evolution of two linear energies during the turbulent cascade in the scalar model with purely derivative self-interactions (assuming $\kappa=0$).}
\label{fig:sat_conv}
\end{figure}

We also analyze the convergence of the solutions shown throughout this work from the linear instability regime through turbulent saturation (assuming $\alpha=0$). In Fig.~\ref{fig:sat_conv}, we compare the behavior of the energy conservation violations in numerical evolutions using different grid spacings of both the scalar evolution scheme and the background spacetime. First, we check explicitly that the growth rate of $E_{11}$ is consistent with that obtained using frequency-domain methods in Ref.~\cite{Siemonsen:2025wib}. As noted above, at sufficiently small $\Delta r_*$, the error of the evolution scheme is dominated by interpolation error from the finite-resolution metric functions $f,l,\Omega$ describing the background spacetime. This is particularly evident during the exponential growth phase in Fig.~\ref{fig:sat_conv}. That is, early on only the $\ell=m=1$ mode is populated with a well-resolved frequency, such that the truncation error of the background dominates the energy conservation violations. Focusing the discussion first on potential-type self-interaction case, as energy is transferred to higher azimuthal modes, for $t/\tau^{\rm EI}_{m=1}\gtrsim 10$, violations of energy conservation are no longer dictated by the resolution of the metric functions, but rather due to the discretization error of the scalar evolution scheme. This can be seen in the central panel of Fig.~\ref{fig:sat_conv}, where the violations decrease with increasing resolution. Higher-order azimuthal modes exhibit higher (spatial and temporal) frequencies. Therefore, as these high-$\ell=m$ modes grow in amplitude, their discretization error begins to dominate the violations of the energy conservation. In the case of purely derivative self-interactions, there exists no conserved energy. Hence, in the bottom panel of Fig.~\ref{fig:sat_conv}, we simply show the consistent convergence of the numerical solution. In this case, energy is injected in the high-$\ell=m$ modes more rapidly compared with the potential-type self-interactions, resulting in larger difference between low- and high-resolutions cases sooner after saturation of the exponential growth. Finally, we checked explicitly that the turbulent cascade results, quoted in the main text, are independent of $\ell_{\max}$, by performing an evolution with $\ell_{\rm max}=31$. 

\end{document}